\documentclass[12pt]{article}
\usepackage{amsmath}

\newcommand{\f}{\frac}
\newcommand{\pdf}[2]{\frac{\partial#1}{\partial#2}}

\newcommand{\intxy}{\int_0^1\!dx\int_0^x\!dy~}
 \newcommand{\Xbar}{\bar X}
\newcommand{\x}[1]{\bar x_{i#1}} \newcommand{\B}[3]{B_#1(#2,#3)}
\newcommand{\bb}[4]{B_{#1#2}(#3,#4)} \newcommand{\C}[2]{C_{#1#2}}
\newcommand{\cc}[5]{C_{#1#2}(#3,#4,#5)} \newcommand{\dd}[2]{D_{#1#2}}
\newcommand{\D}[3]{D_{#1#2#3}} \newcommand{\R}[2]{{\overline R}_{#1#2}}
\newcommand{\dbmoo}{\pdf{\B012}{m_1^2}}
\newcommand{\dbmot}{\pdf{\B023}{m_2^2}}
\newcommand{\dbmof}{\pdf{\B013}{m_1^2}}
\newcommand{\dbmto}{\pdf{\B012}{m_2^2}}
\newcommand{\dbmtt}{\pdf{\B023}{m_3^2}}
\newcommand{\dbmtf}{\pdf{\B013}{m_3^2}}

\begin{document}

\begin{titlepage}
\begin{flushright}
UM-TH-97-07\\ April 1997\\ hep-ph/9704308
\end{flushright}
\vskip 2cm
\begin{center}
{\large\bf Reduction of One-loop Tensor Form-Factors\\ to Scalar
           Integrals:\\ A General Scheme}
\vskip 1cm {\large Ganesh Devaraj\\}
\vskip 1cm and
\vskip 1cm {\large Robin G. Stuart\\}
\vskip 2pt {\it Randall Physics Laboratory, University of Michigan\\ Ann
Arbor, MI 48109-1120, USA}\\
\end{center}
\vskip .5cm

\begin{abstract}
A general method for reducing tensor form factors, that appear in
one-loop calculations in dimensional regularization, to scalar integrals
is presented. The method is an extension of the reduction scheme
introduced by Passarino and Veltman \cite{PassarinoVeltman} and is
applicable in all regions of parameter space including those
where kinematic Gram determinant vanishes. New relations between the
the form factors that valid for vanishing Gram determinant play a key
r\^ole in the extended scheme.
\end{abstract}
\end{titlepage}

\setcounter{footnote}{0} \setcounter{page}{2} \setcounter{section}{0}
\newpage

\section{Introduction}
\label{sec:intro}

The evaluation of one-loop tensor integrals, that was pioneered by Brown
and Feynman \cite{BrownFeynman} and systematized in the work of Passarino
and Veltman \cite{PassarinoVeltman}, is by now a mature science.  The
latter authors defined a set of generic tensor form factors in terms of
which any one-loop tensor integral could be written. Thus, for example, the
3-point tensor integral with two Lorentz indices,
\begin{multline}
C_{\mu\nu}(p_1,p_2;m_1^2,m_2^2,m_3^2)=\\ \int\frac{d^n
q}{i\pi^2}\frac{q_\mu q_\nu}
{[q^2+m_1^2][(q+p_1)^2+m_2^2][(q+p1+p_2)^2+m_3^2]},
\end{multline}
is written in terms of generic tensor form factors
$C_{ij}(p_1^2,p_2^2,p_5^2;m_1^2,m_2^2,m_3^2)$ defined by
\begin{multline}
C_{\mu\nu}(p_1,p_2;m_1^2,m_2^2,m_3^2)=\\ p_{1\mu}p_{1\mu}C_{21}
                    +p_{2\mu}p_{2\nu}C_{22}
                    +\{p_{1\mu}p_{2\nu}+p_{1\nu}p_{2\mu}\}C_{23}
                    +\delta_{\mu\nu}C_{24}
\label{eq:C2defn}
\end{multline}
where $p_5=p_1+p_2$ and we have omitted the arguments of the $C_{ij}$'s in
eq.(\ref{eq:C2defn}).  This choice for defining the $C_{ij}$'s has the
obvious advantage of its great simplicity.  Passarino and Veltman showed
how the $C_{ij}$'s and their analogous 4-point form factors, $D_{ij}$ could
be reduced to scalar integrals and 2-point form factors.

In ref.\cite{LERGI} it was shown how to reduce the $C_{ij}$, $D_{ij}$ to
unique form consisting entirely of scalar integrals.  The advantages of
doing this are manifold.  The resulting expressions are often simpler in
form, containing just a few scalar integrals with coefficients that are
rational functions of the masses and momenta.  All non-trivial analytic
structure, such as cuts in the complex plane, are isolated in the scalar
integrals which can be evaluated separately taking care of these
technicalities \cite{tHooftVeltman79, GeneralD0}.  Having a general and
well-defined procedure for reducing tensor form factors to scalar integrals
lends itself to automation and means that quite lengthy calculations can be
undertaken. In such calculations it is essential that stringent checks be
performed as to the correctness of the final results.  These require the
comparison of two algebraic expressions whose equality can only be
determined if both have been fully reduced to a unique form containing only
scalar integrals with no remaining tensor form factors
\cite{bprimeFCNC,bprimeHiggs}.

Passarino and Veltman applied their reduction scheme numerically to the to
the process $e^+e^-\rightarrow\mu^+\mu^-$ and in doing so highlighted a
significant disadvantage of their approach. It is sometimes the case that
strong numerical cancellations occur during reduction. These authors, for
example, found it necessary to calculate to 40 significant figures in order
to obtain sufficient accuracy. Performing the reduction algebraically as in
ref.\cite{LERGI} has the advantage of eliminating many of the strong
numerical cancellations and hence leads to improved numerical
stability. It can also lead to rather compact and convenient expressions
for physical observables.

Still the reduction scheme of Passarino and Veltman has other vexing
problems. The reduction formulas for 3- and 4-point functions involve the
inversion of kinematic matrices and therefore it breaks down when the
determinants of these matrices, Gram determinants, vanish. In
ref.\cite{LERGI} it was shown how to extend the Passarino-Veltman reduction
scheme to treat situations of vanishing Gram determinants.  This was based
on the observation \cite{GKKS} that in such cases, the $n$-point form
factors reduce to $(n-1)$-point form factors.  This extended reduction
scheme was implemented in the REDUCE program {\tt LERG-I}
\cite{LERGI}. Even here, the procedure broke down for certain
special cases of external momenta and internal masses. In ref.\cite{LERGII}
extensions were made to enlarge the region of parameter space for which the
reduction could be performed but some problematic holes nevertheless
remain. The
extended scheme was implemented in both REDUCE \cite{LERGII} and
Mathematica \cite{LERGIII}. In this paper we present an alternative
reduction scheme for the case of vanishing Gram determinant derived
directly from the original Passarino-Veltman approach. New techniques
involving differentiation with respect to mass arguments are introduced
that make this and earlier schemes viable everywhere in parameter
space, except perhaps where true infrared divergences or mass
singularities appear.

Although expressions that have been algebraically reduced to scalar
integrals exhibit improved numerical stability over those in which the
tensor form factors are calculated numerically in a hierarchical fashion,
they can still become problematic as kinematic determinants become small.
This problem was addressed by van Oldenborgh and Vermaseren
\cite{Vermaseren} who employed a different tensor basis, constructed from
so-called inverse vectors, to define the form factors.  This approach has
the effect of simplifying the contraction of Lorentz indices and yields
expressions that can exhibit improved numerical stability for small Gram
determinants.  Since the form factors that they obtain are linear
combinations of those defined by Passarino and Veltman, their method can
then be viewed as a way of arranging expressions to reduce numerical
instabilities but it does not show, as they state, that the approach of
ref.\cite{LERGI} ``is most suited for numerical application''.

Ezawa {\it et al.}\ \cite{Ezawa} have performed the reduction of the tensor
form factors using an orthonormal tensor basis.

Pittau \cite{Pittau} has proposed a reduction method, based on
$\gamma$-algebra, that avoids the appearance of Gram determinants and is
applicable for loop integrals with at least two massless external legs.

More recently Campbell {\it et al.}\ \cite{CampGlovMill} have pointed out
that the Gram determinants are artifacts of the procedure used to reduce
tensor form factors to scalar integrals.  They show how to write Feynman
parameter representations of tensor integral form factors in terms of
scalar integrals in higher dimensions and derivatives of ordinary scalar
integrals. The various Feynman parameter integrals are separately finite
and highly stable as the condition for vanishing Gram determinant is
approached. In practice, however, evaluating the Feynman parameter
integrals yields terms between which there is eventually a strong additive
cancellation and numerical instabilities do set in although their
appearance is much delayed.

Other tensor integral reduction schemes have been put forward.
In quantum gravity, the computation of tensor integrals by reduction
to scalar integrals was pioneered by Capper and Liebbrandt
\cite{CapLieb} both for the graviton self-energy and the fictitious
particle loop. Davydychev has
\cite{Davydychev} obtained reduction formulas from relations between
integrals in different numbers of dimensions. Recursion relations obtained
recently by Tarasov \cite{Tarasov} have completed the picture to make this
a practical approach but here again problems arise when the Gram
determinant vanishes.  Bern {\it et al.}\cite{BernDixKos1} derive tensor
integrals by differentiation of Feynman parameter representations of scalar
integrals and introduced many of the basic techniques employed in
ref.\cite{CampGlovMill}.

All reduction schemes have certain strengths and weaknesses and may be more
conveniently applied in certain circumstances than others. In the scheme we
present here we take the point of view that one important advantage of
Passarino-Veltman reduction is the simplicity in the way the form-factors
are defined (c.f.\ eq.(\ref{eq:C2defn})). This simplicity may come at the
price of numerical instability near phase space boundaries but even there
appropriate Taylor series expansion of the scalar form factors may
eliminate it. In such cases other methods which employ more numerically
stable form factors may be more easily applied.  The methods given here can
also lead to lengthy algebraic expressions when there are many different
masses and momenta in the problem. These expressions can be costly in terms
of computer storage and processing time although such issues are of
ever-declining importance.

In this work we consider the form factors, $C_{ij}$ and $D_{ij}$, as
well-defined analytic functions of their arguments everywhere in parameter
space. Feynman parameter representations exist for these functions which,
can be expected to be analytic, at least in the physical region, except for
isolated singularities.  The reduction scheme is one way of obtaining an
expression for, or value of, these functions in a given region of parameter
space.  Note that this means, for example, that $C_{21}$ and $C_{22}$ with
arguments $C_{ij}(p^2,p^2,0;m_1^2,m_2^2,m_2^3)$ remain well-defined
entities even though, with $p_1=p_2$, eq.(\ref{eq:C2defn}) does not
uniquely specify them. It also means that relations such as
(\ref{eq:C2defn}) are valid everywhere in the parameter space. This
approach lends itself well to implementations with computer algebra.

Our aim here is to obtain valid analytic expressions for the tensor
form factors in terms of scalar integrals for all regions of the
parameter space where those form factors do not experience genuine
singularities.
We do not address the problem of numerical stability, since it
has been intensely studied elsewhere, but concentrate rather on on the
region where the Gram determinant exactly vanishes. A new ingredient
is the use of
additional relations between tensor integrals and scalar integrals that are
valid only for vanishing Gram determinant. These relations actually follow
directly and straightforwardly from the original reduction formulas given
by Passarino and Veltman.

At first sight it might seem somewhat strange that there is such difficulty
in evaluating the tensor form factors for vanishing Gram determinant. After
all one could, in principle, vary one of the arguments away from region of
vanishing Gram determinant and then use l'H\^opital's rule to take the
limit. One finds, however, that this procedure regenerates a vanishing Gram
determinant in the denominator of the resulting expressions. The new
relations between the form factors allow l'H\^opital's rule to be applied
should it be necessary in extremely singular situations.

\section{Notation}
\label{sec:notation}

In dimensional regularization the dimensionality of space-time, $n$, is
continued away from $n=4$. Quartic, quadratic and logarithmic divergences
correspond to poles at even integer $n\ge 0$, 2 and 4 respectively.  One
introduces the logarithmically divergent constant,

\begin{eqnarray}
\Delta & = & \pi^{\frac{\scriptstyle n}{\scriptstyle 2}-2}
         \Gamma\left(2-\frac{\displaystyle n}{\displaystyle 2}\right)\\ &
         \rightarrow & \frac{-2}{n-4}-\gamma-\ln\pi
\end{eqnarray}
as $n\rightarrow 4$ where $\gamma$ is Euler's constant.

Note that it is essential to keep careful track of background terms that
may be generated when $\Delta$ is multiplied by polynomials in $n$.  For
example
\[
(n-a)\Delta\rightarrow -2+(4-a)\Delta
\]
as $n\rightarrow 4$.

A detailed exposition of dimensional regularization is to be found in
refs~\cite{PassarinoVeltman, Leibbrandt} along with results for certain
standard integrals. For our purposes we need in addition,
\begin{eqnarray}
\int d^np\ \frac{p_\kappa p_\lambda p_\mu p_\nu} {(p^2+2k\cdot
                p+m^2)^\alpha}&=& \frac{i\pi^{\frac{\scriptstyle
                n}{\scriptstyle 2}}} {(m^2-k^2)^{\alpha-\frac{n}{2}}}
                \frac{1}{\Gamma(\alpha)}
                \bigg\{\Gamma\left(\alpha-\frac{\displaystyle
                n}{\displaystyle 2}\right) k_\kappa k_\lambda k_\mu k_\nu
                \nonumber\\
                &+&\frac{1}{2}\Gamma\left(\alpha-1-\frac{\displaystyle n}
                {\displaystyle 2}\right)(m^2-k^2)
                \{kk\delta\}_{\kappa\lambda\mu\nu}\\
                &+&\frac{1}{4}\Gamma\left(\alpha-2-\frac{\displaystyle
                n}{\displaystyle 2} \right) (m^2-k^2)^2
                \{\delta\delta\}_{\kappa\lambda\mu\nu}\bigg\rbrace
                \nonumber
\end{eqnarray}

In defining tensor integrals we shall adopt the notation of
ref.\cite{PassarinoVeltman} in which braces denote symmetrization with
respect to Lorentz indices. Thus for example,

\begin{eqnarray}
\{p\delta\}_{\mu\nu\alpha}&=&p_\mu\delta_{\nu\alpha}
+p_\nu\delta_{\mu\alpha}+p_\alpha\delta_{\mu\nu}\\
\{\delta\delta\}_{\mu\nu\alpha\beta}&=&
\delta_{\mu\nu}\delta_{\alpha\beta}+\delta_{\mu\alpha}\delta_{\nu\beta}
+\delta_{\mu\beta}\delta_{\nu\alpha}
\end{eqnarray}

We will use the notation $B_0(i,j)$ and $C_0(i,j,k)$ in context to denote
the two- and three-point scalar integrals constructed from the $i$-th,
$j$-th and $k$-th denominators of some tensor integral.

\section{One- and two-point form factors}
\label{sec:AandB}

Following Passarino and Veltman \cite{PassarinoVeltman} we use a metric in
which the squares of time-like momenta are negative.

The one-point scalar integral, $A(m^2)$, is quadratically divergent and
defined as,
\begin{eqnarray}
A(m^2)&=&\int \frac{d^nq}{i\pi^2}\frac{1}{q^2+m^2}\\
      &=&\frac{\pi^{\frac{\scriptstyle n}{\scriptstyle 2}-2}}
      {(m^2)^{1-\frac{\scriptstyle n}{\scriptstyle 2}}}
      \Gamma\left(1-\frac{\displaystyle n}{\displaystyle 2}\right)
\end{eqnarray}

The two-point tensor form factors, $B_{ij}(p^2;m_1^2,m_2^2)$, are defined
in Appendix A. These form factors are not independent of one another and
may all be reduced to the form,
\begin{eqnarray}
B_{ij}(p^2;m_1^2,m_2^2)&\rightarrow&\alpha_1 A(m_2^2) +\beta_{12}
                        B_0(p^2;m_1^2,m_2^2)\nonumber\\ & &\quad +\beta_1
                        B_0(0;m_1^2,m_1^2) +\beta_2
                        B_0(0;m_2^2,m_2^2)+\beta_0 \label{eq:Breduct}
\end{eqnarray}
in which $\alpha_1$ and the $\beta$'s are rational functions of the momenta
and masses, $p^2$, $m_1^2$, $m_2^2$,
\begin{equation}
B_0(p^2;m_1^2,m_2^2)=\int \frac{d^nq}{i\pi^2} \frac{1}{\lbrack
  q^2+m_1^2\rbrack\lbrack(q+p)^2+m_2^2\rbrack}
\label{eq:B0def}
\end{equation}
and
\begin{equation}
B_0(0;m^2,m^2)=\Delta-\ln m^2
\label{eq:B00}
\end{equation}

The coefficient $\alpha_1$ is non-zero only for quadratically divergent
$B_{ij}$.  The form factor $B_{43}$ is quartically divergent and requires
the introduction of an additional term but we do not make use of it
explicitly in the the present work.  In addition, the partial derivatives
of any form factor $B_{ij}$ with respect to any of its three arguments can
be written in the form (\ref{eq:Breduct}).

By performing the usual Feynman parameterization of the integral
(\ref{eq:B0def}) one obtains,
\begin{eqnarray}
B_{i1}(p^2;m_1^2,m_2^2)&=& (-)^i\pi^{\frac{\scriptstyle n}{\scriptstyle
2}-2} \Gamma(2-\frac{\displaystyle n}{\displaystyle 2})\int_0^1dz\ z^i
\lbrack az^2+bz+c\rbrack^{\frac{\scriptstyle n} {\scriptstyle 2}-2}
\label{eq:GeneralBi1int1}\\
&=&(-)^i\int_0^1 dz\ z^i\lbrack \Delta-\ln(az^2+bz+c)\rbrack
\label{eq:Bi1int1}
\end{eqnarray}
where $B_{01}\equiv B_0$, $B_{11}\equiv B_1$ and $a=-p^2$,
$b=p^2-m_1^2+m_2^2$, $c=m_1^2-i\epsilon$.  Eq.(\ref{eq:Bi1int1}) follows
from eq.(\ref{eq:GeneralBi1int1}) by dropping terms of ${\cal O}(n-4)$ and
higher. Such terms will be discarded without further comment in what
follows.  In the region of $n=4$ these integrals become
\begin{equation}
B_{i1}(p^2;m_1^2,m_2^2)=\frac{(-)^i}{i+1}\bigg\lbrace\Delta-\ln(a+b+c)
            +\int_0^1 dz\frac{z^{i+1}(2az+b)}{az^2+bz+c}\bigg\rbrace
\label{eq:Bi1int2}
\end{equation}
The integrand can be decomposed as,
\begin{equation}
\frac{z^{i+1}(2az+b)}{az^2+bz+c}= \sum_{j=0}^i\alpha_j z^j+\frac{(\beta_1
z+\beta_0)(2az+b)}{az^2+bz+c}
\end{equation}

Carrying out the integration for all but the term proportional to $\beta_1$
and comparing with (\ref{eq:B00}) and (\ref{eq:Bi1int2}) for $i=0$ yields,

\begin{eqnarray}
B_{i1}(p^2;m_1^2,m_2^2)&=&
\frac{(-)^i}{i+1}\bigg\{\beta_1B_0(p^2;m_1^2,m_2^2)
+\beta_0B_0(0;m_1^2,m_1^2)\nonumber\\ & &
+(1-\beta_0-\beta_1)B_0(0,m_2^2,m_2^2)
+\sum_{j=0}^i\frac{1}{j+1}\alpha_j\bigg\}
\label{eq:Bi1reduct2}
\end{eqnarray}
Full expressions for the form factors $B_{ij}$ can be found in
ref.\cite{LERGI} and will not be repeated here.

The partial derivative of the $B_{i1}(p^2;m_1^2,m_2^2)$ with respect to
$p^2$ form factors arise in practice in wavefunction renormalization
factors and those with respect to the mass arguments are needed in
intermediate steps in the reduction to scalar integrals at kinematic
boundaries. Because of the reduction (\ref{eq:Breduct}), it follows that
one only needs expressions for the partial derivatives of $B_0$ and it
turns out that the partial derivatives themselves take the form
(\ref{eq:Breduct}). From eq.(\ref{eq:Bi1int2}) one has for the region of
$n=4$,
\begin{eqnarray}
\frac{\partial B_0}{\partial
p^2}(p^2;m_1^2,m_2^2)&=& -\int_0^1dz \frac{z(1-z)}{az^2+bz+c}
\label{eq:partialB0partialp2}\\ \frac{\partial B_0}{\partial
m_1^2}(p^2;m_1^2,m_2^2)&=& -\int_0^1dz \frac{(1-z)}{az^2+bz+c}
\label{eq:partialB0partialm2} \end{eqnarray}
The partial derivative with respect to $m_2^2$ easily follows since $B_0$
is symmetric in its mass arguments.

Writing,
\begin{equation}
\frac{z(z-1)}{az^2+bz+c}=\alpha_0+\beta_1\frac{z(2az+b)}{az^2+bz+c}
+\beta_0\frac{(2az+b)}{az^2+bz+c}
\end{equation}
in (\ref{eq:partialB0partialp2}) for which,
\[
\alpha_0=-\frac{b}{a}\frac{2a+b}{b^2-4ac},\ \ \ \ \ \ \
\beta_1=\frac{1}{a}\frac{b^2-2ac+ab}{b^2-4ac},\ \ \ \ \ \ \
\beta_0=\frac{c}{a}\frac{2a+b}{b^2-4ac}
\]
and
\begin{eqnarray}
\frac{1-z}{az^2+bz+c}&=&\alpha_0+\beta_1\frac{z(2az+b)}{az^2+bz+c}
+\beta_0\frac{(2az+b)}{az^2+bz+c}
\end{eqnarray}
in (\ref{eq:partialB0partialm2}) for which,
\[
\alpha_0=\frac{2(2a+b)}{b^2-4ac},\ \ \ \ \ \ \
\beta_1=-\frac{b+2c}{b^2-4ac},\ \ \ \ \ \ \ \beta_0=-\frac{2a+b}{b^2-4ac}
\]
the integrals expressions for the partial derivatives are easily reduced to
the form (\ref{eq:Breduct}).  Full expressions for $\alpha_0$, $\beta_1$,
$\beta_0$ and $-(\beta_1+\beta_0)$ in terms of $p^2$, $m_1^2$, $m_2^2$ are
to be found in Appendix A.

\section{Three-point form factors}
\label{sec:C}

The 3-point scalar form factor $C_0(p_1^2,p_2^2,p_5^2;m_1^2,m_2^2,m_3^2)$
is a function of three external momenta squared and three internal masses.
Following ref.\cite{PassarinoVeltman} we define,
\begin{align}
C_0(p_1^2,p_2^2,p_5^2;&m_1^2,m_2^2,m_3^2)=\nonumber\\
&\int\frac{d^nq}{i\pi^2} \frac{1} {\lbrack q^2+m_1^2\rbrack\lbrack
(q+p_1)^2+m_2^2\rbrack \lbrack (q+p_1+p_2)^2+m_3^2\rbrack}
\label{eq:C0def}
\end{align}
with $p_5=p_1+p_2$.

This may be written as the Feynman parameter integral,
\begin{align}
C_0&(p_1^2,p_2^2,p_5^2;m_1^2,m_2^2,m_3^2)=\nonumber\\
&2\int_0^1dz_1\int_0^1dz_2\int_0^1dz_3\delta(1-z_1-z_2-z_3)\nonumber\\
&\times\int\frac{d^nq}{i\pi^2} \frac{1} {\lbrack
q^2+p_1^2z_1z_2+p_2^2z_1z_3+p_5^2z_1z_3+m_1^2z_1+m_2^2z_2+m_3^2z_3 \rbrack}
\end{align}
that clearly displays the symmetry properties of the arguments and from
which it may be shown that,
\begin{eqnarray}
C_0(p_1^2,p_2^2,p_5^2;m_1^2,m_2^2,m_3^2),\ \ \ \ \
C_0(p_5^2,p_2^2,p_1^2;m_1^2,m_3^2,m_2^2),\nonumber\\
C_0(p_1^2,p_5^2,p_2^2;m_2^2,m_1^2,m_3^2),\ \ \ \ \
C_0(p_2^2,p_5^2,p_1^2;m_2^2,m_3^2,m_1^2),\label{eq:C0symmetries}\\
C_0(p_5^2,p_1^2,p_2^2;m_3^2,m_1^2,m_2^2),\ \ \ \ \
C_0(p_2^2,p_1^2,p_5^2;m_3^2,m_2^2,m_1^2),\nonumber
\end{eqnarray}
are all equal.

In a similar way as for the two-point form factors,
$B_{ij}(p^2,m_1^2,m_2^2)$, one can define a sequence of three-point tensor
form factors, $C_{ij}(p_1^2,p_2^2,p_5^2;m_1^2,m_2^2,m_3^2)$ as shown in
ref.\cite{PassarinoVeltman}.  Passarino and Veltman \cite{PassarinoVeltman}
have shown how to reduce to expressions involving $B_{ij}$ and lower
$C_{ij}$ form factors.  For certain of the tensor three-point form factors,
this method yields two apparently distinct expressions. In
ref.\cite{PassarinoVeltman}, these expressions were evaluated numerically
and their agreement was used as a powerful check of the approach. The
method that they employ can be made use of to algebraically reduce tensor
three-point form factors to the two- and three-point scalar integrals.
\begin{align}
C_{ij}(p_1^2,p_2^2,p_5^2;m_1^2,m_2^2,m_3^2)\rightarrow&
\alpha_1C_0(p_1^2,p_2^2,p_5^2;m_1^2,m_2^2,m_3^2)\nonumber\\
+&\beta_{12}B_0(1,2)+\beta_{13}B_0(1,3)+\beta_{23}B_0(2,3)\nonumber\\
+&\beta_1B_0(0;m_1^2,m_1^2)+\beta_2B_0(0;m_2^2,m_2^2)\nonumber\\
+&\beta_3B_0(0;m_3^2,m_3^2)+\beta_0 \label{eq:Creduct}
\end{align}
where $\alpha_1$ and the $\beta$'s are again rational functions of the
masses and momenta.  The method of ref.\cite{PassarinoVeltman} involves
inverting the matrix,
\begin{equation}
X=\left(
\begin{array}{cc}
p_1^2&p_1\cdot p_2\\ p_1\cdot p_2&p_2^2
\end{array}
\right)
\end{equation}
and consequently breaks down with the vanishing of the determinant at
kinematic boundaries. We will now specify the properties of the external
momenta, $p_1$, $p_2$ and $p_5$ for which the determinant vanishes, that is
when
\begin{equation}
  p_1^2p_2^2-(p_1\cdot p_2)^2=-\frac{1}{4}(p_1^4+p_2^4+p_5^4
                             -2p_1^2p_2^2-2p_1^2p_5^2-2p_2^2p_5^2) =0
\end{equation}
If $p_1=(\vec p_1,iE_1)$, $p_2=(\vec p_2,iE_2)$ and $\theta$ is the angle
between their 3-vector components then,
\begin{equation}
  \cos\theta=\biggl(\frac{E_1}{|\vec p_1|}\biggr) \biggl(\frac{E_2}{|\vec
             p_2|}\biggr) \pm\sqrt{\bigg\lbrack 1-\biggl(\frac{E_1}{|\vec
             p_1|}\biggr)^2\bigg\rbrack \bigg\lbrack
             1-\biggl(\frac{E_2}{|\vec p_2|}\biggr)^2\bigg\rbrack}
\end{equation}
{}From the requirement that $\cos\theta$ be real and $|\cos\theta|\le 1$ it
may be deduced that,
\begin{itemize}
\item If one of the momenta $p_1$, $p_2$, $p_5$ is time--like then the
       other two are as well.  Moreover $p_1=\alpha p_2$ for some real
       constant $\alpha$.
\item If two of the momenta are light--like then the third is as well.
       Moreover $p_1=\alpha p_2$ for some real constant $\alpha$.
\item If one of the momenta is space--like then the other two are either
       both space--like or one is space--like and the other is light--like.
       In this case $p_1$ and $p_2$ can be strictly linearly independent.
\end{itemize}

\section{The Derivative of $C_0$}
\label{sec:dC0}

In the reduction scheme presented in this paper, it is sometimes necessary
to obtain the expression for the derivative of $C_0$ with respect to one of
its mass arguments.  For completeness the expressions for the derivative of
$C_0$ with respect to any of its six arguments are derived in this section.

The scalar integral $C_0$ has a Feynman parameter representation given by
\begin{multline}
C_0(p_1^2,p_2^2,p_5^2,m_1^2,m_2^2,m_3^2) = \intxy \f{1}{a x^2+b y^2+c x y+d
x+e y+f-i\epsilon} \\ {} \label{eq:C0xy}
\end{multline}
where,
\[ \begin{array}{lll}
a=-p_2^2, & b=-p_1^2, & c=p_1^2+p_2^2-p_5^2, \\ d=m_2^2-m_3^2+p_2^2, &
e=m_1^2-m_2^2+p_5^2-p_2^2, & f=m_3^2.
\end{array} \]
The derivative of $C_0$ with respect to any of it's six arguments is,
\begin{equation}
C^\prime_0= \int_0^1\!dx \int_0^x\!dy~\f{\alpha_{x^2} x^2 + \alpha_{y^2}
y^2 + \alpha_{xy} x y + \alpha_x x + \alpha_y y + \alpha_1}{(a x^2 + b y^2
+ c x y + d x + e y + f-i\epsilon)^2}
\label{eq:dC0_general}
\end{equation}
where the coefficients, $\alpha$, are given by
\begin{center}
\begin{tabular}{|c|c|c|c|c|c|c|}  \hline
Derivative w.r.t. & $\alpha_{x^2}$ & $\alpha_{y^2}$ & $\alpha_{xy}$ &
$\alpha_x$ & $\alpha_y$ & $\alpha_1$ \\ \hline $p_1^2$ & 0 & 1 & -1 & 0 & 0
& 0 \\ $p_2^2$ & 1 & 0 & -1 & -1 & 1 & 0 \\ $p_5^2$ & 0 & 0 & 1 & 0 & -1 &
0 \\ $m_1^2$ & 0 & 0 & 0 & 0 & -1 & 0 \\ $m_2^2$ & 0 & 0 & 0 & -1 & 1 & 0
\\ $m_3^2$ & 0 & 0 & 0 & 1 & 0 & -1 \\ \hline
\end{tabular}
\end{center}
It is convenient to define as basis set integrals, $I_i$, for the
decomposition of $C^\prime_0$ as follows
\begin{alignat}{3}
I_1 &\equiv \intxy \f{y(2by+cx+e)}{D^2} & \qquad &I_2 \equiv \intxy
\f{y(2ax+cy+d)}{D^2} \nonumber \\ I_3 &\equiv \intxy \f{x(2by+cx+e)}{D^2} &
\qquad &I_4 \equiv \intxy \f{2ax+cy+d}{D^2} \nonumber \\ I_5 &\equiv \intxy
\f{2by+cx+e}{D^2} & \qquad &I_6 \equiv \intxy \f{dx+ey+2f}{D^2} \nonumber
\end{alignat}
where $D=a x^2+b y^2+c x y +d x+e y+f-i\epsilon$.
It is easy to show that the $I_i$'s may be written in terms of $C_0$ and
the derivatives of $B_0$'s.
\begin{eqnarray*}
I_1 &=& C_0 + \dbmof \\ I_2 &=& \dbmoo - \dbmof \\ I_3 &=& \dbmof - \dbmot
\\ I_4 &=& \dbmoo + \dbmto - \dbmof - \dbmtf \\ I_5 &=& \dbmof + \dbmtf -
\dbmot - \dbmtt \\ I_6 &=& - \dbmoo - \dbmto
\end{eqnarray*}
The derivative of $B_0$ may be expressed in terms of $B_0$'s
as shown in eq.(\ref{eq:dB0m}).
Note however that this equation will yield an ill-defined
$B_0(0;0,0)$ if $m_1^2$ is set equal to zero. Such terms will
generally cancel without any special treatment in expressions for physical
quantities. It may, in cases where infrared divergences or mass
singularities are present, be convenient to carry a non-zero $m_1$
through intermediate steps and set it to zero only in the final results.
With the above relations in hand the reduction of
$C^\prime_0$ is straightforward.
\begin{equation}
C^\prime_0 = \beta_1 I_1 + \beta_2 I_2 + \beta_3 I_3 + \beta_4 I_4 +
 \beta_5 I_5 + \beta_6 I_6
\end{equation}
The equation for the $\beta_i$'s is obtained by equating the coefficients
of $x^2$, $y^2$, etc., in the above.
\begin{equation}
\left(\begin{array}{cccccc} 0&0&c&0&0&0\\2b&c&0&0&0&0\\c&2a&
2b&0&0&0\\0&0&e&2a&c&d\\e&d&0&c&2b&e\\ 0&0&0&d&e&2f \end{array} \right)
\left(\begin{array}{c} \beta_1\\ \beta_2\\ \beta_3\\ \beta_4\\ \beta_5\\
\beta_6 \end{array} \right) = \left(\begin{array}{c} \alpha_{x^2} \\
\alpha_{y^2} \\ \alpha_{xy} \\ \alpha_x \\ \alpha_y \\ \alpha_1 \end{array}
\right)
\label{eq:beta}
\end{equation}
This matrix equation may be solved by first determining $\beta_1$,
$\beta_2$, and $\beta_3$ from the relation
\begin{equation}
\left( \begin{array}{ccc} 0&0&c\\2b&c&0\\c&2a&2b \end{array} \right) \left(
\begin{array}{c} \beta_1\\ \beta_2\\ \beta_3
\end{array} \right) = \left( \begin{array}{c} \alpha_{x^2} \\ \alpha_{y^2}
\\ \alpha_{xy} \end{array} \right)
\end{equation}
and then obtaining $\beta_4$, $\beta_5$, and $\beta_6$ using,
\begin{equation}
\left( \begin{array}{ccc} 2a&c&d\\c&2b&e\\d&e&2f \end{array} \right) \left(
\begin{array}{c} \beta_4\\ \beta_5\\ \beta_6 \end{array} \right) =
\left(\begin{array}{c} \alpha_x \\ \alpha_y \\ \alpha_1
\end{array} \right) - \left( \begin{array}{ccc} 0&0&e\\e&d&0\\0&0&0
\end{array} \right) \left(\begin{array}{c} \beta_1 \\ \beta_2 \\
\beta_3 \end{array} \right)
\label{eq:dC0m}
\end{equation}

It should be noted that the derivative of $C_0$ with respect to any of its
mass arguments reduces to $B_0$'s.  In this case, $\alpha_{x^2} =
\alpha_{y^2}= \alpha_{xy}=0$, and consequently $\beta_1=\beta_2=\beta_3=0$.
Since $I_4$, $I_5$, and $I_6$ can be expressed in terms of just $B_0$'s, it
follows that $\partial C_0(p_1^2,p_2^2,p_5^2,m_1^2,m_2^2,m_3^2)/\partial
m_i^2$ can also be expressed in terms of $B_0$'s.

The expressions for the $\beta_i$'s, for the particular case of the
derivative with respect to $p_1^2$, are given below.
\begin{equation}
\left(\begin{array}{c} \beta_1 \\ \beta_2 \\ \beta_3 \end{array} \right) =
\f{1}{4ab-c^2} \left(\begin{array}{c} 2a+c \\ -2b-c \\ 0 \end{array}
\right)
\end{equation}
\begin{equation}
\left(\begin{array}{c} \beta_4 \\ \beta_5 \\ \beta_6 \end{array} \right) =
\f{2bd+cd-2ae-ce}{2(4ab-c^2)(-bd^2+cde-ae^2+4abf-c^2f)}
\left(\begin{array}{c} de-2cf \\ 4af-d^2 \\ cd-2ae \end{array} \right)
\end{equation}
The expressions for the derivatives with respect to the other two momentum
arguments may be obtained from the symmetry properties of $C_0$
(eq.(\ref{eq:C0symmetries})).

The expressions for the $\beta_i$'s for $\partial C_0/\partial m_1^2$ are,
$\beta_1=\beta_2=\beta_3=0$ and,
\begin{equation}
\left(\begin{array}{c} \beta_4 \\ \beta_5 \\ \beta_6 \end{array} \right) =
\f{-1}{-bd^2+cde-ae^2+4abf-c^2f} \left(\begin{array}{c} de-2cf \\ 4af-d^2
\\ cd-2ae \end{array} \right)
\end{equation}
Again the derivatives with respect to the other masses may be obtained
using the symmetry properties of $C_0$ (eq.(\ref{eq:C0symmetries})).

In case the matrix that needs to be inverted to obtain $\partial
C_0/\partial m_i^2$ (eq.(\ref{eq:dC0m})), has zero determinant, the simple
solution is to obtain the expression for the derivative with one of the one
of the masses shifted away from its value, and since the resulting
expression will contain only $B_0$'s, l'H\^{o}pital's rule
or Taylor series expansion may be applied
to obtain the limit.  Recall that the derivative of $C_0$ with the most
general set of arguments, with respect to any of its mass arguments is
expressible in terms of $B_0$'s.  The expressions for $\partial
B_0/\partial m_i^2$ are given in Appendix A.

\section{Reduction of $C_{ij}$ Form Factors}
\label{sec:rC}

The Passarino-Veltman scheme for the reduction of all $C_{ij}$ form factors
to the basic scalar integrals $C_0$ and $B_0$ breaks down when the
kinematic or Gram determinant, vanishes.  The Gram determinant will be
denoted, ${\cal D}$, and when it vanishes all the $C_{ij}$ form factors,
including the scalar integral $C_0$, can be reduced to linear combinations
of $B_0$'s as long as they are at most logarithmically
divergent. Expressions for the quadratically divergent $C_{ij}$'s contain
the quadratically divergent scalar integral, $A$.

The reduction for $C_0$ is obtained as follows.  The Passarino-Veltman
formula for reducing the scalar integrals $C_{11}$ and $C_{12}$ to $C_0$'s
and $B_0$'s in the notation of ref.\cite{PassarinoVeltman} is,
\begin{equation}
\left(\begin{array}{c} C_{11} \\ C_{12} \end{array} \right) = X^{-1} \left(
      \begin{array}{c} R_1 \\ R_2 \end{array} \right) = \f{1}{\cal D}\left(
      \begin{array}{cc} p_2^2 &-{p_1\!\cdot\!p_2} \\ -{p_1\!\cdot\!p_2}
      &p_1^2 \end{array} \right) \left(\begin{array}{c} R_1 \\ R_2
      \end{array} \right)
\label{eq:C1}
\end{equation}
where,
\begin{eqnarray*}
R_1&=&\frac{1}{2}\{f_1 C_0 + B_0(1,3) - B_0(2,3) \}\\
R_2&=&\frac{1}{2}\{f_2 C_0 + B_0(1,2) - B_0(1,3) \}\\ f_1&=&m_1^2 - m_2^2 -
p_1^2\\ f_2&=&m_2^2 - m_3^2 + p_1^2 - p_5^2
\end{eqnarray*}
and ${\cal D}$ is the determinant of the matrix $X$,
\begin{equation}
{\cal D} = p_1^2 p_2^2 - (p_1\!\cdot\!p_2)^2.
\label{eq:D3}
\end{equation}
When ${\cal D}=0$, in order for $C_{11}$ and $C_{12}$ to exist, the
following equations must be satisfied,
\begin{eqnarray}
p_2^2 ~ R_1 - {p_1\!\cdot\!p_2} ~ R_2 &=& 0 \\ -{p_1\!\cdot\!p_2} ~ R_1 +
p_1^2 ~ R_2 &=& 0
\label{eq:forC0}
\end{eqnarray}
Substituting for $R_1$ and $R_2$ one obtains,
\begin{eqnarray}
C_0 &=& \f{(p_1\!\cdot\!p_2) B_0(1,2) - ({p_1\!\cdot\!p_2} + p_2^2)
        B_0(1,3) + p_2^2 B_0(2,3)}{p_2^2 f_1 - (p_1\!\cdot\!p_2) f_2}
\label{eq:C0pbar1} \\
        C_0 &=& \f{p_1^2 B_0(1,2) - (p_1^2 + {p_1\!\cdot\!p_2}) B_0(1,3) +
        (p_1\!\cdot\!p_2) B_0(2,3)}{(p_1\!\cdot\!p_2) f_1 - p_1^2 f_2}
\label{eq:C0pbar2}
\end{eqnarray}
respectively.  Equations (\ref{eq:C0pbar1}) and (\ref{eq:C0pbar2}) give
explicit expressions for $C_0$ when the kinematic determinant vanishes.
Although both of these expressions are equivalent, it is useful to keep
both expressions at hand because one encounters situations where the
denominator of one of the expressions vanishes but that of the other does
not.  For example, when $p_2^2 = 0$ and $p_1\!\cdot\!p_2 = 0$, the
denominator of the expression in (\ref{eq:C0pbar1}) vanishes but that of
(\ref{eq:C0pbar2}) need not.

The eq.s (\ref{eq:C0pbar1}) and (\ref{eq:C0pbar2}) are new formulas
valid when ${\cal D}=0$ and represent an alternative to the method
described in ref.s\cite{LERGI,LERGII}.

At this point we introduce some notation to make the equations more
concise.  The cofactor matrix of the matrix $X$, will be called $\Xbar$.
Thus,
\[ \Xbar = \left( \begin{array}{cc} \bar x_{11} &\bar x_{12} \\ \bar x_{21}
   &\bar x_{22} \end{array} \right) = \left( \begin{array}{cc} p_2^2
   &-{p_1\!\cdot\!p_2} \\ -{p_1\!\cdot\!p_2} &p_1^2 \end{array} \right) \]
We also define,
\begin{eqnarray}
Y_1&=&\bar x_{11} f_1 + \bar x_{12} f_2 \label{eq:Y1}\\
Y_2&=&\bar x_{21} f_1 + \bar x_{22} f_2 \label{eq:Y2}
\end{eqnarray}
In this notation the two equations for $C_0$, (\ref{eq:C0pbar1}) and
(\ref{eq:C0pbar2}), can be written concisely as,
\begin{equation}
C_0 = \f{1}{Y_i} \{- \bar x_{i2} B_0(1,2) + (\bar x_{i2} - \bar x_{i1})
       B_0(1,3) + \bar x_{i1} B_0(2,3) \}
\label{eq:C0bar}
\end{equation}
where $i = 1$, 2. These eq.s(\ref{eq:C0bar}) run into difficulties
when both of the $Y_i$ vanish. In what follows it will be shown how to
bypass such problems by differentiation with respect to mass arguments.
It can be demonstrated that the vanishing of both $Y_i$'s is a necessary
condition for the breakdown of the method described in ref.\cite{LERGII}
and many of the techniques we introduce here can be used to extend
that method as well.

To obtain the reductions for $C_{11}$ and $C_{12}$ when
${\cal D} = 0$, one could attempt to obtain the limit of the rational
functions in eq.(\ref{eq:C1}) using l'H\^{o}pital's rule but we find that
when the partial derivative of $C_0$ with respect to one of its momentum
arguments, is expressed in terms of $C_0$ and the $B_0$'s, the Gram
determinant, eq.(\ref{eq:D3}), appears in the denominator and the
application of l'H\^{o}pital's rule in this manner does not yield a limit.
Instead, the reductions for $C_{11}$ and $C_{12}$ when ${\cal D} = 0$ can
be obtained by looking at the general formulas for the $C_{2j}$ form
factors,
\begin{eqnarray*}
 \left( \begin{array}{c} C_{21} \\ C_{23} \end{array} \right) = X^{-1}
 \left( \begin{array}{c} R_3 \\ R_5 \end{array} \right) &=& \f{\Xbar} {\cal
 D} \left(\begin{array}{l} \f{1}{2} \{f_1 C_{11} + B_1(1,3) + B_0(2,3) \} -
 C_{24} \\ \f{1}{2} \{f_2 C_{11} + B_1(1,2) - B_1(1,3) \} \end{array}
 \right) \\ \left( \begin{array}{c} C_{23} \\ C_{22} \end{array} \right) =
 X^{-1} \left( \begin{array}{c} R_4 \\ R_6 \end{array} \right) &=&
 \f{\Xbar} {\cal D} \left( \begin{array}{l} \f{1}{2} \{f_1 C_{12} +
 B_1(1,3) - B_1(2,3) \} \\ \f{1}{2} \{f_2 C_{12} - B_1(1,3) \} - C_{24}
 \end{array} \right)
\end{eqnarray*}
Hence when ${\cal D} = 0$, we must have,
\begin{eqnarray}
\bar x_{i1} R_3 + \bar x_{i2} R_5 &=& 0 \\ \bar x_{i1} R_4 + \bar x_{i2}
R_6 &=& 0
\end{eqnarray}
Substituting for $R_3$, $R_4$, $R_5$, and $R_6$ one obtains,
\begin{eqnarray}
Y_i C_{11} &=& - \bar x_{i2} B_1(1,2) + (\bar x_{i2} - \bar x_{i1})
               B_1(1,3) - \bar x_{i1} B_0(2,3) + 2 \bar x_{i1} C_{24} \quad
               \quad \label{eq:C1barpre1} \\ Y_i C_{12} &=& (\bar x_{i2} -
               \bar x_{i1}) B_1(1,3)
               + \bar x_{i1} B_1(2,3) + 2 \bar x_{i2} C_{24}
\label{eq:C1barpre2}
\end{eqnarray}
Unlike the case for $C_0$ where the right hand side did not include any
$C_{ij}$ form factors, here the logarithmically divergent $C_{24}$ is
encountered on the right hand side.  The appearance of $C_{24}$ causes no
difficulties because it is given in closed form by,
\begin{equation}
C_{24} = \f{1}{4} - \f{1}{4}\{f_1 C_{11} + f_2 C_{12} + 2 m_1^2 C_0 -
             B_0(2,3)\}
\label{eq:C24}
\end{equation}
Note that this is not affected by a vanishing denominator when ${\cal
D}=0$.

Substituting this expression for $C_{24}$ into (\ref{eq:C1barpre1}) and
(\ref{eq:C1barpre2}) one
obtains,
\begin{equation}
\left(\begin{array}{cc}\x1f_1+2Y_i&\x1f_2\\ \x2f_1&2Y_i+\x2f_2\end{array}
\right) \left( \begin{array}{c}\C11\\ \C12\end{array}\right) =
\left(\begin{array}{c}\R11\\ \R12\end{array}\right)
\label{eq:C1bar}
\end{equation}
where,
\begin{eqnarray*}
\R11 &=& - 2 m_1^2 \bar x_{i1} C_0 - 2 \bar x_{i2} B_1(1,2) + 2 (\bar
          x_{i2} - \bar x_{i1}) B_1(1,3) \hspace{1.0in}\\ & & - \bar x_{i1}
          B_0(2,3) + \bar x_{i1} \\ \R12 &=& - 2 m_1^2 \bar x_{i2} C_0 + 2
          (\bar x_{i2} - \bar x_{i1}) B_1(1,3) + 2 \bar x_{i1} B_1(2,3) \\
          & & + \bar x_{i2} B_0(2,3) + \bar x_{i2}
\end{eqnarray*}
Once the substitution for $C_0$ is made using eq.(\ref{eq:C0bar})
and $B_1$ is
written in terms of $B_0$ \cite{LERGI}, the expressions for $C_{11}$ and
$C_{12}$ will be given purely in terms of $B_0$'s.

The formulas for the reduction for $C_{2j}$, $C_{3j}$, and $C_{4j}$ form
factors, when ${\cal D} = 0$, are obtained in a similar manner from the
general formulas for the reduction of the $C_{3j}$, $C_{4j}$, and $C_{5j}$
form factors respectively.  The difference between the $C_{2j}$ and higher
form factors, and the ones treated above, $C_0$ and $C_{1j}$, is that the
former have form factors that multiply tensors that have one or more
$\delta_{\mu\nu}$.  All the form factors that multiply tensors containing
one or more factors of $\delta_{\mu\nu}$, have formulas that do not suffer
from the problem of a vanishing denominator.  These form factors are
$C_{24}$, $C_{35}$, $C_{36}$, $C_{46}$, $C_{47}$, $C_{48}$, and $C_{49}$.
The formula for the reduction of $C_{24}$ is given above in
eq.(\ref{eq:C24}).  The formulas for $C_{35}$, $C_{36}$, $C_{46}$, etc. are
given in Appendix B.  The rest of the form factors, those that multiply
tensors without any $\delta_{\mu \nu}$'s in them, are given by formulas
similar to (\ref{eq:C1bar}) but involving larger matrices. The form factors
$C_{21}$, $C_{22}$, and $C_{23}$ are given by a set of three simultaneous
linear equations, $C_{31}$ to $C_{34}$ by a set of four equations, and
$C_{41}$ to $C_{45}$ by a set of five equations.  The relevant formulas
appear in Appendix~C.

The reduction formulas for these form factors have denominators that are
proportional to powers of $Y_i$.  As discussed earlier, when only one of
the $Y_i$'s is equal to zero no problem arises but when both $Y_1$ and
$Y_2$ are equal to zero, the limit of these rational functions has to be
obtained using l'H\^{o}pital's rule.  We discuss two separate cases here:
\begin{itemize}
\item At least one of the $\bar x_{ij}$ is not zero : In this case the
limit is easily obtained using l'H{\^o}pital's rule or
Taylor series expansion
where the differentiation is performed with respect to one of the mass
arguments.
\item All of the $\bar x_{ij}$ are zero :
This condition arises if and only if
all three of the external momenta squared are equal to zero.  In this case
the $Y_i$ cannot be shifted away from zero by shifting one of the mass
arguments.  The reduction formulas for this case are obtained as follows.
The derivation for $C_0$ is given below.  The rest of the form factors are
obtained in the same manner.  To obtain the reduction for $C_0$, we go back
to eq.(\ref{eq:C1}).  When $p_1^2 = p_2^2 = p_5^2 = 0$, note that every
element of ``$X^{-1}$'' is of the $0/0$ form.  Applying l'H{\^o}pital's
rule on every element of $X^{-1}$, using $p_1^2$, $p_2^2$, and $p_5^2$ as
independent variables and differentiating with respect to $p_5^2$,
eq.(\ref{eq:C1}) becomes,
\begin{equation}
\left(\begin{array}{c} C_{11} \\ C_{12} \end{array} \right) = \f{-1}{p_1 \!
      \cdot \! p_2} \left( \begin{array}{cc} 0 & \f{1}{2} \\ \f{1}{2} & 0
      \end{array} \right) \left(\begin{array}{c} R_1 \\ R_2 \end{array}
      \right)
\label{eq:C1zero}
\end{equation}
Since $p_1 \! \cdot \! p_2=0$ and $C_{11}$ and $C_{12}$ are assumed to
exist we must have that $R_1 = 0$ and $R_2 = 0$. Substituting for $R_1$ and
$R_2$ one obtains,
\begin{equation}
C_0 = \frac{1}{f_1} \{B_0(2,3) - B_0(1,3)\} = \frac{1}{f_2} \{B_0(1,3) -
      B_0(1,2)\}
\label{eq:C0zerobar}
\end{equation}
Substituting $p_1^2=p_2^2=p_5^2=0$ in (\ref{eq:C0zerobar}) one obtains,
\begin{eqnarray}
C_0(0,0,0,m_1^2,m_2^2,m_3^2) &=& \f{B_0(0,m_2^2,m_3^2) -
B_0(0,m_1^2,m_3^2)}{m_1^2 - m_2^2} \\ &=& \f{B_0(0,m_1^2,m_3^2) -
B_0(0,m_1^2,m_2^2)}{m_2^2 - m_3^2}
\end{eqnarray}
If in the above formulas the denominator vanishes, the limit is obtained by
differentiating with respect to one of the masses.
\end{itemize}

\subsection{Example}

The 1-loop QED contribution to the anomalous magnetic moment ($g-2$) of
the electron was first calculated by Schwinger \cite{Schwinger}.
Because of the many singularities involved this is one of the most
pathological cases that can be encountered.
The result can be derived in terms of 3-point form factors to be
\begin{equation}
\frac{g-2}{2}=-\left(\frac{me}{2\pi}\right)^2\left(C_{11}+C_{12}\right)
\end{equation}
in which the arguments of the form factors are
$C_{ij}(-m^2,0,-m^2;\lambda^2,m^2,m^2)$ where $m$ is the electron mass
and $e$ is its charge. Although absent in the final result, infrared
divergences may appear in intermediate steps and are regulated by giving
the photon a mass, $\lambda$. The given set of momenta leads to a
vanishing Gram determinant and for the given masses causes the extended
method, described in ref.\cite{LERGII}, to break down. In the method
described here the denominators $Y_i$ in eq.s (\ref{eq:Y1}) and
(\ref{eq:Y2}) both vanish. To alleviate this $m_3$ can be shifted
away from $m$ and the reduction performed in the usual way. l'H\^opital's
or Taylor series expansion is then used to take the limit
$m_3\rightarrow m$.
This same approach works when applied to the method of ref.\cite{LERGII}
and involves differentiation with respect to the mass arguments of
$B_0$ by means of the formulas given in Appendix A. The photon mass,
$\lambda$ can then be safely set to zero. Note that the form factors
$C_{12}$, $C_{22}$ and $C_{23}$ all blow up as $m_3\rightarrow m$,
even with a finite photon mass, but that none of these occur in the
expression for the physical quantity $(g-2)$. Following this procedure
one obtains
\begin{equation}
C_{11}+C_{12}=\frac{1}{2m^2}
              \left[1+B_0(0;m^2,m^2)-B_0(-m^2;0,m^2)\right]
\end{equation}
using the result
\begin{equation}
B_0(-m^2;0,m^2)=B_0(0;m^2,m^2)+2
\end{equation}
this gives
\[ \frac{g-2}{2} = \frac{\alpha}{2\pi} \]
in agreement with the well-known result.

\section{Reduction of $D_{ij}$ Form Factors}
\label{sec:rD}

The derivation of the reduction formulas for the $D_{ij}$ form factors when
the kinematic determinant, which we again call $\cal D$, vanishes, is
obtained in an identical manner as for the $C_{ij}$ form factor reductions.
The kinematic determinant, $\cal D$, is given by,
\[ {\cal D} = p_1^2 p_2^2 p_3^2 - p_1^2 (p_2\!\cdot\!p_3)^2 - p_2^2
   (p_1\!\cdot\!p_3)^2 - p_3^2 (p_1\!\cdot\!p_2)^2 +
   2(p_1\!\cdot\!p_2)(p_1\!\cdot\!p_3)(p_2\!\cdot\!p_3) \] To obtain the
   formula for $D_0$ when ${\cal D}=0$, we start with the general formula
   for the reduction of the $D_{1j}$ form factors.
\begin{equation}
\left(\begin{array}{c} D_{11} \\ D_{12} \\ D_{13} \end{array} \right) =
\f{1}{\cal D} \left( \begin{array}{ccc} \bar x_{11} & \bar x_{12} & \bar
x_{13} \\ \bar x_{21} & \bar x_{22} & \bar x_{23} \\ \bar x_{31} & \bar
x_{32} & \bar x_{33} \end{array} \right) \left(\begin{array}{c} R_1 \\ R_2
\\ R_3 \end{array} \right)
\label{eq:D1}
\end{equation}
where,
\begin{eqnarray}
R_1&=&\{f_1 D_0 + C_0(1,3,4) - C_0(2,3,4)\}/2\\ R_2&=&\{f_2 D_0 +
C_0(1,2,4) - C_0(1,3,4)\}/2\\ R_3&=&\{f_3 D_0 + C_0(1,2,3) - C_0(1,2,4)\}/2
\end{eqnarray}
\[
f_1 = m_1^2 - m_2^2 - p_1^2~, ~~~~f_2 = m_2^2 - m_3^2 + p_1^2 - p_5^2~,
   ~~~~f_3 = m_3^2 - m_4^2 - p_4^2 + p_5^2~. \] The elements of the
   co-factor matrix, $\bar X$, are,
\begin{equation}
\left( \begin{array}{ccc} \bar x_{11} & \bar x_{12} & \bar x_{13} \\ \bar
x_{21} & \bar x_{22} & \bar x_{23} \\ \bar x_{31} & \bar x_{32} & \bar
x_{33} \end{array} \right) = \left( \begin{array}{ccc} p_2^2 p_3^2 -
p_{23}^2 & p_{13} p_{23} - p_{12} p_3^2 & p_{12} p_{23} - p_2^2 p_{13} \\
p_{13} p_{23} - p_{12} p_3^2 & p_1^2 p_3^2 - p_{13}^2 & p_{12} p_{13} -
p_1^2 p_{23} \\ p_{12} p_{23} - p_2^2 p_{13} & p_{12} p_{13} - p_1^2 p_{23}
& p_1^2 p_2^2 - p_{12}^2 \end{array} \right) \quad
\label{eq:Xbar}
\end{equation}
where $p_{ij}$ is used to represent $p_i \cdot p_j$.  Using the same
argument as in section \ref{sec:rC}, when ${\cal D}=0$, in order for
$D_{11}$, $D_{12}$, and $D_{13}$ to exist, the following relation must
hold.
\begin{multline}
D_0 = \f{1}{Y_i} \{ - \bar x_{i3} C_0(1,2,3) + (\bar x_{i3} - \bar x_{i2})
 C_0(1,2,4) \\ + (\bar x_{i2} - \bar x_{i1}) C_0(1,3,4) + \bar x_{i1}
 C_0(2,3,4)\}
\label{eq:D0bar}
\end{multline}
where,
\[ Y_i = \bar x_{i1} f_1 + \bar x_{i2} f_2 + \bar x_{i3} f_3 \]
and $i$ can take on the values 1, 2 or 3.

The reduction formulas for the $D_{1j}$ and higher form factors when ${\cal
D} = 0$ are obtained as in the case of the $C_{ij}$ form factors.  Here
again the $D_{ij}$ form factors that multiply tensors containing one or
more $\delta_{\mu \nu}$'s, are given by formulas which do not suffer from a
vanishing denominator for any set of momenta or masses.  The rest of the
form factors are obtained by solving simultaneous linear equations.  These
form factors have denominators which are proportional to powers of $Y_i$.

As in the case of the $C_{ij}$ form factors, when all of the $Y_i$ are
zero, the limit of the rational functions are obtained using
l'H\^{o}pital's rule or Taylor series expansion.
Again there are two cases to consider:
\begin{itemize}
\item At least one of the $\bar x_{ij}$ are non-zero: The limit is obtained
using l'H{\^o}pital's rule where the differentiation is performed with
respect to one of the mass arguments.
\item All the $\bar x_{ij}$ are zero: This case is treated as for the
$C_{ij}$ form factors, but there is one difference to note.  Recall that in
the case of the $C_{ij}$ form factors, all the $\bar x_{ij}$ vanished if
and only if all the external momenta squared were zero, but here all the
$\bar x_{ij}$ could vanish even when none of the external momenta squared
are zero. An example of this is when all the external momenta are
proportional to one another.  Again we use the same technique described in
the earlier section.  Apply l'H{\^o}pital's rule to each element of
``$X^{-1}$''.  Using $p_1^2$, $p_2^2$, $p_3^2$, $p_{12}$, $p_{13}$, and
$p_{23}$ as independent variables and differentiating with respect to
$p_1^2$ one obtains,
\begin{equation}
\f{\bar X}{\cal D} \rightarrow \f{1}{p_2^2 p_3^2 - p_{23}^2} \left(
\begin{array}{ccc} 0 & 0 & 0 \\ 0 & p_3^2 & - p_{23} \\ 0 & - p_{23} &
p_2^2 \end{array} \right)
\label{eq:Xbarprime}
\end{equation}
If the relation (\ref{eq:Xbarprime}) is again of the $0/0$ form, ie.  if
$p_2^2 = p_3^2 = p_{23} = 0$, then the differentiation is performed again,
but this time with respect to, say $p_2^2$, to obtain,
\begin{equation}
\f{\bar X}{\cal D} \rightarrow \f{1}{p_3^2} \left( \begin{array} {ccc} 0 &
0 & 0 \\ 0 & 0 & 0 \\ 0 & 0 & 1 \end{array} \right)
\label{eq:Xbar2prime}
\end{equation}
Proceeding with the assumption that (\ref{eq:Xbarprime}) is not of the
$0/0$ form, eq.(\ref{eq:D1}) becomes,
\begin{equation}
\left(\begin{array}{c} D_{11} \\ D_{12} \\ D_{13} \end{array} \right) =
\f{1}{p_2^2 p_3^2 - p_{23}^2} \left( \begin{array}{ccc} 0 & 0 & 0 \\ 0 &
p_3^2 & - p_{23} \\ 0 & - p_{23} & p_2^2 \end{array} \right)
\left(\begin{array}{c} R_1 \\ R_2 \\ R_3 \end{array} \right)
\end{equation}
Since all the $\bar x_{ij}$ are zero, $p_2^2 p_3^2 - p_{23}^2 = 0$ and
therefore we must have,
\[ p_3^2 R_2 - p_{23} R_3 = 0 ~~~~~~~~{\rm and}~~~~~~~~p_{23} R_2 - p_2^2
R_3  = 0 \]
Substituting for $R_2$ and $R_3$ one obtains the expression for
$D_0$.  If (\ref{eq:Xbarprime}) is of the $0/0$ form, then using
(\ref{eq:Xbar2prime}) we get $R_3 = 0$, and hence the expression for $D_0$.
The expression obtained for $D_0$ will be in terms of $C_0$, but these
$C_0$'s necessarily reduce to $B_0$'s.  That is, the vanishing of all the
$\bar x_{ij}$ is a sufficient condition for the reduction of $D_0$, and
hence all $D$ form factors, to $B_0$'s.
\end{itemize}

\section{Alternate Derivation of the Reduction Formulas}
\label{sec:alt}

The formulas obtained above for the reduction of the $C_{ij}$ and $D_{ij}$
form factors when their respective kinematic determinants vanish, may be
obtained in another manner described below.

Consider the derivation of the reduction formulas for the $C_{ij}$ form
factors.  The reduction for $C_0$ is obtained as before but for the $C_1$
form factors, the derivation proceeds in an identical manner as in section
\ref{sec:rC} only up to eq.(\ref{eq:C1barpre2}). The $C_{24}$ form factor
can be handled differently.  Instead of using eq.(\ref{eq:C24}), one can
use the relation,
\begin{equation}
C_{24} = \f{1}{Y_i}\{-\bar x_{i2}B_{22}(1,2)+(\bar x_{i2}-\bar x_{i1})
B_{22}(1,3)+\bar x_{i1}B_{22}(2,3)\}
\label{eq:C24bar}
\end{equation}
Eq.(\ref{eq:C24bar}) is only valid when ${\cal D}=0$, and it is derived
from the general formulas for the reduction of $C_{35}$ and $C_{36}$ form
factors, given by \cite{PassarinoVeltman},
\begin{equation}
\left(\begin{array}{c} C_{35}\\C_{36}\end{array}\right)=\f{\bar X}{\cal D}
\left(\begin{array}{c}\f{1}{2}\{f_1C_{24}+B_{22}(1,3)-B_{22}(2,3)\}\\
\f{1}{2}\{f_2C_{24}+B_{22}(1,2)-B_{22}(1,3)\}\end{array}\right)
\label{eq:C35C36}
\end{equation}
When (\ref{eq:C24bar}) is substituted for $C_{24}$ in
eq.s (\ref{eq:C1barpre1}) and (\ref{eq:C1barpre2}), the formulas for
$C_{11}$ and $C_{12}$ are obtained
in terms of $B_{ij}$ form factors which can be eventually reduced to
$B_0$'s.

The reduction of the $D_{ij}$ form factors proceeds in the same manner.
The advantage of this method is that one obtains formulas that are not
coupled.  That is, one does not have to solve simultaneous equations to
obtain formulas for the individual form factors, as was necessary in the
previous method.  The disadvantage here becomes evident when this method is
used for obtaining the formulas for the higher form factors.  Note that to
obtain the formulas for the $C_{1j}$ form factors we had to use the general
formulas for the reduction of $C_{3j}$ form factors.  In the formula for
the reduction of the $C_{2j}$ form factors, $C_{35}$ and $C_{36}$ show up.
{}From the general formulas for the $C_{4j}$ form factors one obtains a
formula for $C_{35}$ and $C_{36}$ in terms of $B_{ij}$ form factors and the
quadratically divergent $C_{49}$.  Then from the general formulas for
$C_{5j}$ form factors one can obtain a formula for $C_{49}$ in terms of
$B_{ij}$ form factors alone.  Specifically the $B_{ij}$ form factors that
appear in the formula for $C_{49}$ is the quartically divergent form
factor, $B_{43}$.  That is, the quadratically divergent $C_{49}$ is
obtained as the difference of quartically divergent $B_{43}$.  Substituting
this formula for $C_{49}$ into the formulas for the $C_{35}$ and $C_{36}$
form factors, gives $C_{35}$ and $C_{36}$ purely in terms of $B_{ij}$ form
factors which can then be substituted into the formulas for the $C_{2j}$
form factors to give $C_{21}$, $C_{22}$ and $C_{23}$ in terms of $B_{ij}$
form factors alone.  Similarly, to obtain the formulas for the $C_{3j}$
form factors one has to drive general formulas for the $C_{7j}$ form
factors, which contain the highly divergent $B_{64}$. The situation is even
worse for the $C_{4j}$ form factors and similar problems are faced in the
case of the $D_{ij}$ form factors.

Therefore the previous derivation is simpler since to obtain the formulas
for $C_{ij}$ form factors, we only have to look at the general formulas of
the $C_{i+1,j}$ form factors, and likewise for the $D_{ij}$ form factors.
Also, one does not have to deal with the highly divergent form factors
which show up in this alternate derivation.  But no matter how the
reduction formulas are derived, once the formulas are reduced to $B_0$'s
and $C_0$'s, they take on unique forms.

\section{Comparison with Earlier Work}
\label{sec:comp}

The method employed in ref.s\cite{LERGI,LERGII,LERGIII} to treat the
reduction of scalar integrals when the Gram determinant vanished yielded
spurious logarithmic divergences that were observed to cancel in physical
results. Thus, for example, the logarithmic divergences that appeared in
the expressions for $C_{11}$ and $C_{12}$, canceled in practical situations
when the $C_{\mu ...}$ were constructed and the indices were contracted
with the
appropriate Lorentz tensors.  In this section we will show how the
cancellation occurs, list all the situations under which the cancellation
necessarily occurs and also give the situation under which this
cancellation might not occur.

The formula given for the $C_{1j}$ form factors in Ref~\cite{LERGI} is,
\begin{eqnarray}
C_{11}&=&\alpha_{12}B_1(1,2)+\alpha_{13}B_1(1,3)-\alpha_{23}B_0(2,3)
\label{eq:C1lerg1} \\
C_{12}&=&\alpha_{13}B_1(1,3)+\alpha_{23}B_1(2,3)
\label{eq:C1lerg2}
\end{eqnarray}
where the $\alpha_{ij}$ are,
\[ \alpha_{12}=-\bar x_{22}/Y_2,~~~~\alpha_{13}
=(\bar x_{22}-\bar x_{21})/Y_2,~~~\alpha_{23}=\bar x_{21}/Y_2 \] Comparing
eq.s (\ref{eq:C1lerg1}) and (\ref{eq:C1lerg2}) with the complete
expressions given in
eq.s (\ref{eq:C1barpre1}) and (\ref{eq:C1barpre2}), it is evident that the
terms involving the
logarithmically divergent $C_{24}$ are missing from the expressions in
eq.s (\ref{eq:C1lerg1}) and (\ref{eq:C1lerg2}).

The reason why the terms involving $C_{24}$ cancel in practice is shown
below.  If one constructs $C_{\mu}=p_{1\mu}C_{11}+p_{2\mu}C_{12}$ using
$C_{11}$ and $C_{12}$ from eq.s (\ref{eq:C1barpre1}) and
(\ref{eq:C1barpre2}), then one obtains,
\begin{multline}
C_{\mu}=\f{1}{Y_i}\big(-p_{1\mu}\bar x_{i2}B_1(1,2)+(p_{1\mu}+p_{2\mu})
(\bar x_{i2}-\bar x_{i1})B_1(1,3) \\ +p_{2\mu}\bar
x_{i1}B_1(2,3)-p_{1\mu}\bar x_{i1}B_0(2,3)+ 2(p_{1\mu}\bar
x_{i1}+p_{2\mu}\bar x_{i2})C_{24}\big)
\end{multline}
Note that the term involving $C_{24}$ is proportional to the light-like
vector
\begin{equation}
P_{\mu}=p_{1\mu}\bar x_{i1}+p_{2\mu}\bar x_{i2}
\end{equation}
We note that for all higher rank tensor integrals, $C_{\mu\nu}$,
$C_{\mu\nu\lambda}$, etc., the terms that are missing from the expressions
in ref.\cite{LERGI} are always proportional to the vector $P_{\alpha}$,
where $\alpha$ is one of the indices of the tensor integral.

Remembering that at the outset we have assumed ${\cal D}=0$, we note the
following properties of the vector $P_{\mu}$
\begin{itemize}
\item $P_{\mu}=0$~~if~~$p_1=\alpha p_2$~~for some real $\alpha$.
\item $p_{1\mu}\!\cdot\!P_{\mu}=0$.
\item $p_{2\mu}\!\cdot\!P_{\mu}=0$.
\end{itemize}
It was stated in section \ref{sec:C} that when ${\cal D}=0$, it is
necessarily true that $p_1=\alpha p_2$, unless at least two of the three
external momenta, were space--like.  Therefore, as long as the external
momenta are not space--like, $P_{\mu}$ is identically zero.  If at least
two of the external momenta are space--like, $P_{\mu}$ is not identically
zero, but if it is dotted with any of the external momenta or any linear
combination of them, it vanishes. Under either of these conditions, which
is almost always the case, the term involving $C_{24}$ vanishes.  The only
condition under which this term can survive is if at least two of the
external momenta are space--like {\it and\/} it is contracted with a vector
which is linearly independent of the external momenta of the three point
function.  Such conditions may occur for the process
$e^+e^-\rightarrow\nu_e\bar\nu_e\gamma$ produced by $t$-channel exchange of
of a $W$-boson. There the one-loop corrections to the $W^+W^-\gamma$ vertex
can yield a form-factor having two external space-like momenta (i.e.\ those
of the $W$'s) and one external light-like momentum (i.e.\ that of the
external photon).  In such a case the method used in
ref.s\cite{LERGI,LERGII,LERGIII} might yield incorrect results but would be
indicated by the non-cancellation of the extra spurious diverges.  In the
program {\tt LERG-I} this would also be indicated by the appearance of
arbitrary constants in physical results.

In the case of the $D_{ij}$ form factors, the situation is similar.  The
``missing terms'' are always proportional to the vector $p_{1\mu}\bar
x_{i1}+p_{2\mu}\bar x_{i2}+p_{3\mu}\bar x_{i3}$.  The properties of this
vector are identical to the one above.  It is identically zero if any two
of $p_{1\mu}$, $p_{2\mu}$, and $p_{3\mu}$ are linearly dependent, and it
vanishes if it is dotted with any of its external momenta, and hence any
linear combination of them.  Here again, the only condition under which the
``missing terms'' can survive, is if all three external momenta,
$p_{1\mu}$, $p_{2\mu}$, and $p_{3\mu}$, are linearly independent {\it and}
if $D_{\mu ...}$ is not contracted with any vector which is linearly
dependent with the external momenta of the four-point tensor integral.

\section{Summary and Conclusion}

We have extended the method for the reduction of Lorentz tensor form
factors to scalar integrals, introduced by Passarino and Veltman, to
regions of parameter space including those of vanishing Gram
determinants. No attempt has been made to handle
numerical instabilities that may arise and the region of vanishing Gram
determinant is approached since this problem as been amply discussed in the
literature. If required the methods described here can be used in this
critical region by employing a Taylor series expansion.

The technique presented here may also be applied in the reduction of
five-point and higher functions \cite{NeervenVermaseren} when their Gram
determinants vanish.  This is necessarily the case for six-point and higher
functions in four space-time dimensions.

In the foregoing it is implicitly assumed that the calculations are
performed in a covariant gauge. In axial gauge, for example, new
denominators appear in the Feynman integrals that are not immediately
amenable to the methods described here.

Once again it should be emphasized that many of the formulas that have been
derived in this paper are valid only for the divergent and finite parts
of the integrals and do not necessarily apply to the terms of
${\cal O}(n-4)$ and higher. Thus the results cannot be used for the
insertion of subdiagrams to form 2-loop or higher diagrams.

The reduction of tensor form factors to scalar integrals can also be
implemented at the 2-loop level \cite{WeigScharfBohm} but new
complications arise. Whereas at 1-loop there is just a single topology
for each of the 2-, 3- and 4-point scalar integrals, this is no longer
true at 2-loops. For the 2-point function at 2-loops there are 20
possible topologies for the scalar integrals. Some of these can be
calculated in closed form \cite{ScharfTausk} but it is known that general
case is not expressible in terms of polylogarithms and so are probably
best calculated numerically.

\setcounter{section}{0} \def\thesection{Appendix \Alph{section}}
\def\theequation{\Alph{section}.\arabic{equation}}

\setcounter{equation}{0}
\section{}

The 2-point form factors, $B_{ij}$, are defined following Passarino and
Veltman \cite{PassarinoVeltman} via the relations
\begin{equation*}
B_0;B_{\mu};B_{\mu\nu}; B_{\mu\nu\alpha};B_{\mu\nu\alpha\beta}(p;m_1,m_2)
=\int\frac{d^nq}{i\pi^2}\frac{1;q_\mu;q_\mu q_\nu;q_\mu q_\nu q_\alpha;
q_\mu q_\nu q_\alpha q_\beta} {[q^2+m_1^2][(q+p)^2+m_2^2]}
\end{equation*}

By invoking Lorentz covariance one may define the form factor,
\[B_\mu=p_\mu B_1\]
Here and in what follows the arguments of the $B_{ij}$'s are taken to be
$B_{ij}(p^2;m_1^2,m_2^2)$.  Similarly,
\begin{eqnarray*}
B_{\mu\nu}&=&p_\mu p_\nu B_{21}+\delta_{\mu\nu}B_{22}\\
B_{\mu\nu\alpha}&=&p_\mu p_\nu p_\alpha B_{31}
+\{p\delta\}_{\mu\nu\alpha}B_{32}\\ B_{\mu\nu\alpha\beta}&=&p_\mu p_\nu
p_\alpha p_\beta B_{41} +\{pp\delta\}_{\mu\nu\alpha\beta}B_{42}
+\{\delta\delta\}_{\mu\nu\alpha\beta}B_{43}
\end{eqnarray*}
The braces are used to denote the symmetrized product in the Lorentz
indices.  The form factors $B_{i1}$ (we take $B_0\equiv B_{01}$, $B_1\equiv
B_{11}$) are all logarithmically divergent as signaled by their having
poles at even integer $n\ge 4$.

$A$ and $B_{i2}$ are quadratically divergent with poles at even integer
$n\ge 2$.  $B_{43}$ is quartically divergent with poles at even integer
$n\ge0$.

Expressions for the derivative of $B_0$ with respect to its momentum
argument can be found in ref.\cite{LERGI} in the form (\ref{eq:Breduct})
and various special cases and recursion relations are to be found in
ref.\cite{LERGII}.  Here we give the expressions for the derivatives with
respect to both the momentum and mass arguments. Neither derivative
exists at the physical threshold, $p^2=-(m_1+m_2)^2$.

Away from threshold
\begin{eqnarray}
\frac{\partial B_0}{\partial p^2}(p^2;m_1^2,m_2^2) &=&
-\frac{p^2m_1^2+p^2m_2^2+m_1^4-2m_1^2m_2^2+m_2^4}{p^2D}
B_0(p^2;m_1^2,m_2^2) \nonumber\\ &
&+{m_1^2}\frac{(p^2-m_2^2+m_1^2)}{p^2D}B_0(0;m_1^2,m_1^2)\nonumber\\ &
&+{m_2^2}\frac{(p^2-m_1^2+m_2^2)}{p^2D}B_0(0;m_2^2,m_2^2)\\ &
&-\frac{(p^2-m_1^2+m_2^2)(p^2-m_2^2+m_1^2)}{p^2D}\nonumber
\end{eqnarray}
provided the denominator $p^2D\ne0$. Here
\begin{eqnarray}
D&=&(p^4+m_1^4+m_2^4+2p^2m_1^2+2p^2m_2^2-2m_1^2m_2^2) \\ &=&\lbrack
 p^2+(m_1+m_2)^2\rbrack\lbrack p^2+(m_1-m_2)^2\rbrack.
\end{eqnarray}

For $p^2=0$ and $m_1^2\ne m_2^2$
\begin{multline}
\frac{\partial B_0}{\partial p^2}(p^2;m_1^2,m_2^2)\bigg\vert_{p^2=0}=
-\frac{m_1^2+m_2^2}{2(m_1^2-m_2^2)^2}\\
-\frac{m_1^2m_2^2}{(m_1^2-m_2^2)^3}
[B_0(0;m_1^2,m_1^2)-B_0(0;m_2^2,m_2^2)]
\end{multline}
which reduces to
\begin{equation}
\frac{\partial B_0}{\partial p^2}(p^2;m_1^2,m_2^2)\bigg\vert_{p^2=0}=
-\frac{1}{6m^2}
\end{equation}
when $m^2=m_1^2=m_2^2$.

For the special case, $p^2=-(m_1-m_2)^2$, for which $D$ vanishes
\begin{multline}
\frac{\partial B_0}{\partial p^2}
(p^2;m_1^2,m_2^2)\bigg\vert_{p^2=-(m_1-m_2)^2}\\
=-\frac{2}{ p^2}\left\{
1-\frac{(m_1^2-m_2^2)}{4p^2}
\left\lbrack B_0(0;m_1^2,m_1^2)-B_0(0;m_2^2,m_2^2)\right\rbrack
\right\}\Bigg\vert_{p^2=-(m_1-m_2)^2}.
\end{multline}

Provided $D\ne 0$ the derivative with respect to $m_1^2$ is
\begin{eqnarray}
\frac{\partial B_0}{\partial m_1^2}(p^2;m_1^2,m_2^2) &=&
\frac{p^2+m_1^2-m_2^2}{D}B_0(p^2;m_1^2,m_2^2) \nonumber \\ &&
-\frac{p^2+m_2^2+m_1^2}{D}B_0(0;m_1^2,m_1^2)
+\frac{2m_2^2}{D}B_0(0;m_2^2,m_2^2) \nonumber \\ &&
-{2}\frac{(p^2+m_1^2-m_2^2)}{D}.
\label{eq:dB0m}
\end{eqnarray}
When $p^2=-(m_1-m_2)^2$
\begin{multline}
\frac{\partial B_0}{\partial m_1^2}
(p^2;m_1^2,m_2^2)\bigg\vert_{p^2=-(m_1-m_2)^2}\\
=\frac{1}{m_1(m_1-m_2)}
-\frac{1}{2p^2}
\left\lbrack B_0(0;m_1^2,m_1^2)-B_0(0;m_2^2,m_2^2)\right\rbrack
\Bigg\vert_{p^2=-(m_1-m_2)^2}.
\end{multline}
For $p^2=0$ and $m_1^2\ne m_2^2$ eq.(\ref{eq:dB0m}) becomes
\begin{multline}
\frac{\partial B_0}{\partial m_1^2}
(p^2;m_1^2,m_2^2)\bigg\vert_{p^2=-(m_1-m_2)^2}\\
=-\frac{1}{(m_1^2-m_2^2)}
-\frac{m_2^2}{(m_2^2-m_1^2)^2}
\left\lbrack B_0(0;m_1^2,m_1^2)-B_0(0;m_2^2,m_2^2)\right\rbrack
\end{multline}
which reduces to
\begin{equation}
\frac{\partial B_0}{\partial p^2}(p^2;m_1^2,m_2^2)\bigg\vert_{p^2=0}=
-\frac{1}{2m^2}
\end{equation}
for $m_1^2=m_2^2=m^2$.

The partial derivative with respect to $m_2^2$ can be obtained from the
fact that $B_0$ is symmetric in its mass arguments.

\setcounter{equation}{0}
\section{}

The following are the reduction formulas for the $C_{ij}$ form factors that
multiply tensors which have one $\delta_{\mu\nu}$ in them.  These form
factors are logarithmically divergent.
\begin{eqnarray*}
\C24&=&-\f12\left[-\f12+\f12\{f_1\C11+f_2\C12+2m_1^2C_0-\B023\}\right] \\
\C35&=&-\f13\left[\f13+\f12\{f_1\C21+f_2\C23+2m_1^2\C11+\B023\}\right] \\
\C36&=&-\f13\left[\f16+\f12\{f_1\C23+f_2\C22+2m_1^2\C12-\B123\}\right] \\
\C46&=&-\f14\left[-\f14+\f12\{f_1\C31+f_2\C33+2m_1^2\C21-\B023\}\right] \\
\C47&=&-\f14\left[-\f1{12}+\f12\{f_1\C34+f_2\C32+2m_1^2\C22-\bb2123\}\right]
\\ \C48&=&-\f14\left[-\f18+\f12\{f_1\C33+f_2\C34+2m_1^2\C23+\B123\}\right]
\end{eqnarray*}
All the logarithmically divergent $C_{ij}$ form factors given above can be
written using a single formula, with a change in notation, as follows,
\begin{eqnarray*}
C_{mn1}=\f{-1}{m+n}\biggl[ [C_{mn1}]+\f12\{f_1C_{\{m+1\}n0}
+f_2C_{m\{n+1\}0} \hspace{2.5cm} \\ \hfill
+2m_1^2C_{mn0}+(-1)^{m+1}B_{n0}(2,3)\}\biggr]
\end{eqnarray*}
where,
\[ [C_{mn1}] = \lim_{n\to4}(n-4)C_{mn1} = \f{(-1)^{m+n+1}}{(m+n+2)(n+1)} \]
This notation will become evident with an example.
\[C_{\mu\nu}=p_{1\mu}p_{1\nu}C_{200}+p_{2\mu}p_{2\nu}C_{020}+\{p_1p_2\}
_{\mu\nu}C_{110}+\delta_{\mu\nu}C_{001}\] The first index of $C$ counts the
number of $p_1$'s in the tensor that it multiplies, the second index, the
number of $p_2$'s and the third index, the number of $\delta$'s.  In the
case of the $B_{ij}$ form factors, the first index counts the number of
$p$'s, and the second index, the number of $\delta$'s.

Finally the only quadratically divergent form factor $C_{49}$ is given by,
\begin{eqnarray*}
\C49=-\f14\biggl[\f12f_1\C35+\f12f_2\C36+m_1^2\C24-\bb2223+\f14p_2^2\bb2123
\hspace{0.2cm}\\ \hfill +\f14(p_2^2-m_2^2+m_3^2)\B123-\f14m_2^2\B023 \\
\hfill+\f1{48}(p_1^2-p_2^2+p_5^2+4m_1^2-2m_2^2-2m_3^2)\biggr]
\end{eqnarray*}

The $D_{ij}$ form factors that multiply tensors that have one
$\delta_{\mu\nu}$ in them are given by,
\begin{eqnarray*}
\dd27&=&-1\left[\f12\{f_1\dd11+f_2\dd12+f_3\dd13
 +2m_1^2D_0-C_0(2,3,4)\}\right]\\
 \D311&=&-\f12\left[\f12\{f_1\dd21+f_2\dd24+f_3\dd25
 +2m_1^2\dd11+C_0(2,3,4)\}\right]\\
 \D312&=&-\f12\left[\f12\{f_1\dd24+f_2\dd22+f_3\dd26+2m_1^2\dd12
 -\cc11234\}\right]\\
 \D313&=&-\f12\left[\f12\{f_1\dd25+f_2\dd26+f_3\dd23+2m_1^2\dd13
 -\cc12234\}\right]\\ \D416&=&-\f13\left[\f12\{f_1\dd31+f_2\dd34+f_3\dd35
 +2m_1^2\dd21-C_0(2,3,4)\}\right]\\
 \D417&=&-\f13\left[\f12\{f_1\dd36+f_2\dd32+f_3\dd38+2m_1^2\dd22
 -\cc21234\}\right]\\
 \D418&=&-\f13\left[\f12\{f_1\dd37+f_2\dd39+f_3\dd33+2m_1^2\dd23
 -\cc22234\}\right]\\
 \D419&=&-\f13\left[\f12\{f_1\dd34+f_2\dd36+f_3\D310+2m_1^2\dd24
 +\cc11234\}\right]\\
 \D420&=&-\f13\left[\f12\{f_1\dd35+f_2\D310+f_3\dd37+2m_1^2\dd25
 +\cc12234\}\right]\\
 \D421&=&-\f13\left[\f12\{f_1\D310+f_2\dd38+f_3\dd39+2m_1^2\dd26
 -\cc23234\}\right]
\end{eqnarray*}
This set of $D_{ij}$ form factors given above can be written using a single
formula, with a change in notation, as,
\begin{eqnarray*}
D_{mnu1}=\f{-1}{m+n+u+1}\biggl[\f12\{f_1D_{\{m+1\}nu0}+f_2D_{m\{n+1\}u0}
+f_3D_{mn\{u+1\}0}\\ \hfill
+2m_1^2D_{mnu0}+(-1)^{m+1}C_{nu0}(2,3,4)\}\biggr]
\end{eqnarray*}
The notation is as described in the case of the $C$ form factors.  The
first, second, and third index of the $D_{ij}$ form factor count the number
of $p_1$'s, $p_2$'s, and $p_3$'s respectively, in the tensor that it
multiplies, and the last index counts the number of $\delta$'s.

Finally, the only logarithmically divergent $D_{ij}$ form factor,
$D_{422}$, is given by,
\begin{eqnarray*}
\D422=-\f13\biggl[-\f14+\f12\{p_1^2\D416+p_2^2\D417
 +p_3^2\D418+2(p_1{\cdot}p_2)\D419 \hspace{1.6cm}\\ \hfill
 +2(p_1{\cdot}p_3)\D420+2(p_2{\cdot}p_3)\D421 +m_1^2\dd27-\cc24234\}\biggr]
\end{eqnarray*}

\setcounter{equation}{0}
\section{}

The following are the reduction formulas for the $C_{ij}$ form factors when
${\cal D}=0$.  The formulas for $C_0$ and the $C_{1j}$ form factors are
given in section \ref{sec:rC}.  The formulas for the $C_{ij}$ form factors
that multiply tensors that have one or more $\delta_{\mu\nu}$ are given in
Appendix~B.  The rest of the $C_{ij}$ form factors are given here.

\vspace{12pt}
\noindent
{\bf $C_{2j}$ Form Factors}

\[\left(\begin{array}{ccc}2\x1f_1+3Y_i&2\x1f_2&0\\\x2f_1&4Y_i&
\x1f_2\\0&2\x2
f_1&3Y_i+2\x2f_2\end{array}\right) \left(\begin{array}{c} \C21\\ \C23\\
\C22
\end{array}\right)
= \left(\begin{array}{c}\R21\\ \R23\\ \R22 \end{array}\right)\]
\begin{eqnarray*}
\R21&=&-4m_1^2\x1\C11-3\x2\bb2112+3(\x2-\x1)\bb2113\\ & &
 +\x1\B023-\f43\x1\\
 \R23&=&-2m_1^2(\x2\C11+\x1\C12)+3(\x2-\x1)\bb2113-2\x1\bb1123
 \hspace{0.3in}\\ & & -\x2\B023-\x1/3-2\x2/3\\
 \R22&=&-4m_1^2\x2\C12+3(\x2-\x1)\bb2113+3\x1\bb2123\\ & &
 +2\x2\bb1123-\f23\x2
\end{eqnarray*}

\vspace{12pt}
\noindent
{\bf $C_{3j}$ Form Factors}

\[\left(\begin{array}{cccc} 3\x1f_1+4Y_i&3\x1f_2&0&0\\ \x2f_1&\x1f_1+5Y_i&
2\x1f_2&0\\0&2\x2f_1&5Y_i+\x2f_2&\x1f_2\\0&0&3\x2f_1&4Y_i+3\x2f_2
\end{array} \right) \times {\bf C}_3={\overline{\bf R}}_3\] where,
\[\{{\bf C}_3\}^T=(\begin{array}{cccc}\C31&\C33&\C34&\C32\end{array})
\hspace{0.3in} \{{\overline{\bf R}}_3\}^T
=(\begin{array}{cccc}\R31&\R33&\R34&\R32\end{array})\]
\begin{eqnarray*}
\R31&=&-6m_1^2\x1\C21-4\x2\bb3112+4(\x2-\x1)\bb3113\hspace{0.2in}\\ & &
 -2\x1\B023+\f32\x1\\
 \R33&=&-2m_1^2(\x2\C21+2\x1\C23)+4(\x2-\x1)\bb3113+2\x1\B123\\ & &
 +\x2\B023+\f12\x1+\f12\x2\\
 \R34&=&-2m_1^2(2\x2\C23+\x1\C22)+4(\x2-\x1)\bb3113-3\x1\bb2123
 \hspace{0.2in}\\
 & & -2\x2\B123+\f16\x1+\f12\x2\\
 \R32&=&-6m_1^2\C22+4(\x2-\x1)\bb3113+4\x1\bb3123\\ & &
 +3\x2\bb2123+\f12\x2
\end{eqnarray*}

\newpage
\noindent
{\bf $C_{4j}$ Form Factors}

\begin{eqnarray*}
\left(\begin{array}{ccccc} 4\x1f_1+5Y_i&4\x1f_2&0&0&0\\
\x2f_1&2\x1f_1+6Y_i& 3\x1f_2&0&0\\ 0&2\x2f_1&7Y_i&2\x1f_2&0\\
0&0&3\x2f_1&6Y_i+2\x2f_2&\x1f_2\\
0&0&0&4\x2f_1&5Y_i+4\x2f_2\end{array}\right)~~\\ \hfill \times {\bf
C}_4={\overline{\bf R}}_4
\end{eqnarray*}
where,
\begin{eqnarray*}
\{{\bf C}_4\}^T&=&(\begin{array}{ccccc}\C41&\C43&\C45&\C44&\C42\end{array})
\\ \{{\overline{\bf R}}_4\}^T
&=&(\begin{array}{ccccc}\R41&\R43&\R45&\R44&\R42\end{array})
\end{eqnarray*}
\begin{eqnarray*}
\R41&=&-8m_1^2\x1\C31-5\x2\bb4112+5(\x2-\x1)\bb4113 \\ & &
 +\x1\B023-\f85\x1 \\
 \R43&=&-2m_1^2(\x2\C31+3\x1\C33)+5(\x2-\x1)\bb4113-2\x1\B123 \\ & &
 -\x2\B023-\f35\x1-\f25\x2 \\
 \R45&=&-4m_1^2(\x2\C33+\x1\C34)+5(\x2-\x1)\bb4113+3\x1\bb2123 \\ & &
 +2\x2\B123-\f2{15}(2\x1+3\x2) \\
 \R44&=&-2m_1^2(3\x2\C34+\x1\C32)+5(\x2-\x1)\bb4113-4\x1\bb3123
 \hspace{0.2in}\\ & & -3\x2\bb2123-\f1{10}\x1-\f25\x2 \\
 \R42&=&-8m_1^2\x2\C32+5(\x2-\x1)\bb4113+5\x1\bb4123 \\ & &
 +4\x2\bb3123-\f25\x2
\end{eqnarray*}

\setcounter{equation}{0}
\section{}

The following are the formulas for the $D_{ij}$ form factor reductions when
${\cal D}=0$.  The formula for the reduction of $D_0$ is given in section
\ref{sec:rD}.  The formulas for the reduction of all $D_{ij}$ form factors
which multiply tensors that have one or more $\delta_{\mu\nu}$ in them are
given in Appendix~B.  The formulas for the rest of the $D_{ij}$ form factor
reductions are given below.

\vspace{12pt}
\noindent
{\bf $D_{1j}$ Form Factors}
\[\left(\begin{array}{ccc}Y_i+\x1f_1&\x1f_2&\x1f_3\\\x2f_1&Y_i+\x2f_2&\x2f_3\\
\x3f_1&\x3f_2&Y_i+\x3f_3\end{array}\right)\left(\begin{array}{c}\dd11\\
\dd12\\ \dd13\end{array}\right)=\left(\begin{array}{c}\R11\\ \R12\\ \R13
\end{array}\right)\]
\begin{eqnarray*}
\R11&=&-2m_1^2\x1D_0-\x3\cc11123+(\x3-\x2)\cc11124 \hspace{0.9in}\\
&&+(\x2-\x1)\cc11134\\ \R12&=&-2m_1^2\x2D_0-\x3\cc12123+(\x3-\x2)\cc12124\\
&&+(\x2-\x1)\cc11134+\x1\cc11234+\x2C_0(2,3,4)\\
\R13&=&-2m_1^2\x3D_0+(\x3-\x2)\cc12124+(\x2-\x1)\cc12134\\
&&+\x1\cc12234+\x3C_0(2,3,4)
\end{eqnarray*}

The $D_{2j}$ and higher $D_{ij}$ form factor formulas involve large
matrices which do not fit neatly in the page and so the equations are not
written in the matrix form.

\vspace{12pt}
\noindent
{\bf $D_{2j}$ Form Factors}
\begin{multline}
(Y_i+\x1f_1)\dd21+\x1f_2\dd24+\x1f_3\dd25=\\
-2m_1^2\x1\dd11-\x3\cc21123+(\x3-\x2)\cc21124\\ +(\x2-\x1)\cc21134
\nonumber
\end{multline}
\begin{multline}
\x2f_1\dd21+(3Y_i-\x3f_3)\dd24+\x2f_3\dd25+\x1f_2\dd22+\x1f_3\dd26=\\
-2m_1^2(\x2\dd11+\x1\dd12)+2(\x2-\x1)\cc21134-2\x3\cc23123\\
+2(\x3-\x2)\cc23124-\x1\cc11234-\x2C_0(2,3,4) \nonumber
\end{multline}
\begin{multline}
\x3f_1\dd21+\x3f_2\dd24+(3Y_i-\x2f_2)\dd25+\x1f_2\dd26+\x1f_3\dd23=\\
-2m_1^2(\x3\dd11+\x1\dd13)+2(\x3-\x2)\cc23124\\
+2(\x2-\x1)\cc23134-\x1\cc12234-\x3C_0(2,3,4) \nonumber
\end{multline}
\begin{multline}
\x2f_1\dd24+(Y_i+\x2f_2)\dd22+\x2f_3\dd26=\\
-2m_1^2\x2\dd12+(\x2-\x1)\cc21134+\x1\cc21234\\
-\x3\cc22123+(\x3-\x2)\cc22124+\x2\cc11234 \nonumber
\end{multline}
\begin{multline}
\x3f_1\dd24+\x2f_1\dd25+\x3f_2\dd22+(3Y_i-\x1f_1)\dd26+\x2f_3\dd23=\\
-2m_1^2(\x3\dd12+\x2\dd13)+2(\x3-\x2)\cc22124+2(\x2-\x1)\\
\cc23134+2\x1\cc23234+\x3\cc11234+\x2\cc12234 \nonumber
\end{multline}
\begin{multline}
\x3f_1\dd25+\x3f_2\dd26+(Y_i+\x3f_3)\dd23=\\
-2m_1^2\x3\dd13+(\x3-\x2)\cc22124+(\x2-\x1)\cc22134\\
+\x1\cc22234+\x3\cc12234 \nonumber
\end{multline}

\vspace{12pt}
\noindent
{\bf $D_{3j}$ Form Factors}
\begin{multline}
(Y_i+\x1f_1)\dd31+\x1f_2\dd34+\x1f_3\dd35=\\
-2m_1^2\x1\dd21-\x3\cc31123+(\x3-\x2)\cc31124\\ +(\x2-\x1)\cc31134
\nonumber
\end{multline}
\begin{multline}
\x2f_1\dd31+(3Y_i+2\x1f_1+\x2f_2)\dd34\\ \qquad \quad
+\x2f_3\dd35+2\x1f_2\dd36+2\x1f_3\D310= \hfill \\ \qquad \quad \qquad \quad
-2m_1^2(\x2\dd21+2\x1\dd24)-3\x3\cc33123 \hfill \\ \qquad \quad \qquad
\quad \qquad \quad +3(\x3-\x2)\cc33124+3(\x2-\x1)\cc31134 \hfill \\
+\x1\cc11234+\x2\cc0{}234 \nonumber
\end{multline}
\begin{multline}
\x3f_1\dd31+\x3f_2\dd34\\ \qquad \quad +(3Y_i+2\x1f_1+\x3f_3)\dd35
+2\x1f_2\D310+2\x1f_3\dd37= \hfill \\ \qquad \quad \qquad \quad
-2m_1^2(\x3\dd21+2\x1\dd25)+3(\x3-\x2)\cc33124 \hfill \\
+3(\x2-\x1)\cc33134+\x1\cc12234+\x3\cc0{}234 \nonumber
\end{multline}
\begin{multline}
2\x2f_1\dd34+(4Y_i+\x2f_2-\x3f_3)\dd36\\ \qquad \quad
+2\x2f_3\D310+\x1f_2\dd32+\x1f_3\dd38= \hfill \\ \qquad \quad \qquad \quad
-2m_1^2(2\x2\dd24+\x1\dd22)-3\x3\cc34123\hfill \\ \qquad \quad \qquad \quad
\qquad \quad +3(\x3-\x2)\cc34124+3(\x2-\x1)\cc31134 \hfill \\
-2\x1\cc21234-2\x2\cc11234 \nonumber
\end{multline}
\begin{multline}
\x3f_1\dd34+\x2f_1\dd35+\x3f_2\dd36\\ \qquad \quad
+4Y_i\D310+\x2f_3\dd37+\x1f_2\dd38+\x1f_3\dd39= \hfill \\ \qquad \quad
\qquad \quad -2m_1^2(\x3\dd24+\x2\dd25+\x1\dd26)+3(\x2-\x1)\cc33134 \hfill
\\ \qquad \quad \qquad \quad \qquad \quad
+3(\x3-\x2)\cc34124-2\x1\cc23234\hfill \\ -\x3\cc11234-\x2\cc12234
\nonumber
\end{multline}
\begin{multline}
2\x3f_1\dd35+2\x3f_2\D310\\ \qquad \quad
+(4Y_i-\x2f_2+\x3f_3)\dd37+\x1f_2\dd39+\x1f_3\dd33=\hfill\\
-2m_1^2(2\x3\dd25+\x1\dd23)+3(\x3-\x2)\cc34124\\
+3(\x2-\x1)\cc34134-2\x1\cc22234-2\x3\cc12234 \nonumber
\end{multline}
\begin{multline}
\x2f_1\dd36+(Y_i+\x2f_2)\dd32+\x2f_3\dd38=\\
-2m_1^2\x2\dd22+(\x2-\x1)\cc31134+\x1\cc31234\\
-\x3\cc32123+(\x3-\x2)\cc32124+\x2\cc21234 \nonumber
\end{multline}
\begin{multline}
\x3f_1\dd36+2\x2f_1\D310+\x3f_2\dd32\\ \qquad \quad
+(3Y_i+2\x2f_2+\x3f_3)\dd38+2\x2f_3\dd39=\hfill \\ \qquad \quad \qquad
\quad -2m_1^2(\x3\dd22+2\x2\dd26)+3(\x3-\x2)\cc32124 \hfill \\ \qquad \quad
\qquad \quad \qquad \quad +3(\x2-\x1)\cc33134+3\x1\cc33234\hfill \\
+\x3\cc21234+2\x2\cc23234 \nonumber
\end{multline}
\begin{multline}
2\x3f_1\D310+\x2f_1\dd37+2\x3f_2\dd38\\ \qquad \quad
+(3Y_i+\x2f_2+2\x3f_3)\dd39+\x2f_3\dd33=\hfill\\ \qquad \quad \qquad \quad
-2m_1^2(2\x3\dd26+\x2\dd23)+3(\x3-\x2)\cc32124 \hfill \\ \qquad \quad
\qquad \quad \qquad \quad +3(\x2-\x1)\cc34134+3\x1\cc34234\hfill \\
+\x2\cc22234+2\x3\cc23234 \nonumber
\end{multline}
\begin{multline}
\x3f_1\dd37+\x3f_2\dd39+(Y_i+\x3f_3)\dd33=\\
-2m_1^2\x3\dd23+(\x3-\x2)\cc32124+(\x2-\x1)\cc32134\\
+\x1\cc32234+\x3\cc22234 \nonumber
\end{multline}

\vspace{12pt}
\noindent
{\bf $D_{4j}$ Form Factors}
\begin{multline}
(Y_i+\x1f_1)\dd41+\x1f_2\dd44+\x1f_3\dd45=\\
-2m_1^2\x1\dd31-\x3\cc41123+(\x3-\x2)\cc41124\\ +(\x2-\x1)\cc41134
\nonumber
\end{multline}
\begin{multline}
\x2f_1\dd41+(5Y_i+2\x1f_1-\x3f_3)\dd44\\ \qquad \quad
+\x2f_3\dd45+3\x1f_2\D410+3\x1f_3\D413= \hfill \\ \qquad \quad \qquad \quad
-2m_1^2(\x2\dd31+3\x1\dd34)-4\x3\cc43123 \hfill \\ \qquad \quad \qquad
\quad \qquad \quad +4(\x3-\x2)\cc43124+4(\x2-\x1)\cc41134\hfill\\
-\x1\cc11234-\x2\cc0{}234 \nonumber
\end{multline}
\begin{multline}
\x3f_1\dd41+\x3f_2\dd44\\ \qquad \quad
+(5Y_i+2\x1f_1-\x2f_2)\dd45+3\x1f_2\D413+3\x1f_3\D411=\hfill\\
-2m_1^2(\x3\dd31+3\x1\dd35)+4(\x3-\x2)\cc43124\\
+4(\x2-\x1)\cc43134-\x1\cc12234-\x3\cc0{}234 \nonumber
\end{multline}
\begin{multline}
\x2f_1\dd44+(3Y_i-\x3f_3)\D410+\x2f_3\D413+\x1f_2\dd46+\x1f_3\D414=\\
-2m_1^2(\x2\dd34+\x1\dd36)-2\x3\cc45123+2(\x3-\x2)\cc45124\\
+2(\x2-\x1)\cc41134+\x1\cc21234+\x2\cc11234 \nonumber
\end{multline}
\begin{multline}
\x3f_1\dd44+\x2f_1\dd45+\x3f_2\D410\\ \qquad \quad
+(5Y_i+\x1f_1)\D413+\x2f_3\D411+2\x1f_2\D414+2\x1f_3\D415=\hfill \\ \qquad
\quad \qquad \quad -2m_1^2(\x3\dd34+\x2\dd35+2\x1\D310)\hfill \\ \qquad
\quad \qquad \quad \qquad \quad
+4(\x3-\x2)\cc45124+4(\x2-\x1)\cc43134\hfill \\
+2\x1\cc23234+\x3\cc11234+\x2\cc12234 \nonumber
\end{multline}
\begin{multline}
\x3f_1\dd45+\x3f_2\D413\\ \qquad \quad
+(3Y_i-\x2f_2)\D411+\x1f_2\D415+\x1f_3\dd48=\hfill \\
-2m_1^2(\x3\dd35+\x1\dd37)+2(\x3-\x2)\cc45124\\
+2(\x2-\x1)\cc45134+\x1\cc22234+\x3\cc12234 \nonumber
\end{multline}
\begin{multline}
3\x2f_1\D410+(5Y_i+2\x2f_2-\x3f_3)\dd46\\ \qquad \quad
+3\x2f_3\D414+\x1f_2\dd42+\x1f_3\dd47=\hfill \\ \qquad \quad \qquad \quad
-2m_1^2(3\x2\dd36+\x1\dd32)-4\x3\cc44123\hfill \\ \qquad \quad \qquad \quad
\qquad \quad +4(\x3-\x2)\cc44124+4(\x2-\x1)\cc41134\hfill \\
-3\x1\cc31234-3\x2\cc21234 \nonumber
\end{multline}
\begin{multline}
\x3f_1\D410+2\x2f_1\D413+\x3f_2\dd46\\ \qquad \quad
+(5Y_i+\x2f_2)\D414+2\x2f_3\D415+\x1f_2\dd47+\x1f_3\D412=\hfill \\ \qquad
\quad \qquad \quad -2m_1^2(\x3\dd36+2\x2\D310+\x1\dd38)\hfill \\ \qquad
\quad \qquad \quad \qquad \quad
+4(\x3-\x2)\cc44124+4(\x2-\x1)\cc43134\hfill \\
-3\x1\cc33234-\x3\cc21234-2\x2\cc23234 \nonumber
\end{multline}
\begin{multline}
2\x3f_1\D413+\x2f_1\D411+2\x3f_2\D414\\ \qquad \quad
+(5Y_i+\x3f_3)\D415+\x2f_3\dd48+\x1f_2\D412+\x1f_3\dd49=\hfill \\ \qquad
\quad \qquad \quad -2m_1^2(2\x3\D310+\x2\dd37+\x1\dd39)\hfill \\ \qquad
\quad \qquad \quad \qquad \quad
+4(\x3-\x2)\cc44124+4(\x2-\x1)\cc45134\hfill \\
-3\x1\cc34234-\x2\cc22234-2\x3\cc23234 \nonumber
\end{multline}
\begin{multline}
3\x3f_1\D411+3\x3f_2\D415\\ \qquad \quad
+(5Y_i-\x2f_2+2\x3f_3)\dd48+\x1f_2\dd49+\x1f_3\dd43=\hfill \\
-2m_1^2(3\x3\dd37+\x1\dd33)+4(\x3-\x2)\cc44124\\
+4(\x2-\x1)\cc44134-3\x1\cc32234-3\x3\cc22234 \nonumber
\end{multline}
\begin{multline}
\x2f_1\dd46+(Y_i+\x2f_2)\dd42+\x2f_3\dd47=\\
-2m_1^2\x2\dd32-\x3\cc42123+(\x3-\x2)\cc42124\\
+(\x2-\x1)\cc41134+\x1\cc41234+\x2\cc31234 \nonumber
\end{multline}
\begin{multline}
\x3f_1\dd46+3\x2f_1\D414+\x3f_2\dd42\\ \qquad \quad
+(5Y_i-\x1f_1+2\x2f_2)\dd47+3\x2f_3\D412=\hfill \\ \qquad \quad \qquad
\quad -2m_1^2(\x3\dd32+3\x2\dd38)+4(\x3-\x2)\cc42124 \hfill \\ \qquad \quad
\qquad \quad \qquad \quad +4(\x2-\x1)\cc43134+4\x1\cc43234 \hfill \\
+\x3\cc31234+3\x2\cc33234 \nonumber
\end{multline}
\begin{multline}
\x3f_1\D414+\x2f_1\D415+\x3f_2\dd47\\ \qquad \quad
+(3Y_i-\x1f_1)\D412+\x2f_3\dd49=\hfill \\ \qquad \quad \qquad \quad
-2m_1^2(\x3\dd38+\x2\dd39)+2(\x3-\x2)\cc42124 \hfill \\ \qquad \quad \qquad
\quad \qquad \quad +2(\x2-\x1)\cc45134+2\x1\cc45234\hfill \\
+\x3\cc33234+\x2\cc34234 \nonumber
\end{multline}
\begin{multline}
3\x3f_1\D415+\x2f_1\dd48+3\x3f_2\D412\\ \qquad \quad
+(5Y_i-\x1f_1+2\x3f_3)\dd49+\x2f_3\dd43=\hfill \\ \qquad \quad \qquad \quad
-2m_1^2(3\x3\dd39+\x2\dd33)+4(\x3-\x2)\cc42124\hfill \\ \qquad \quad \qquad
\quad \qquad \quad +4(\x2-\x1)\cc44134+4\x1\cc44234 \hfill \\
+\x2\cc32234+3\x3\cc34234 \nonumber
\end{multline}
\begin{multline}
\x3f_1\dd48+\x3f_2\dd49+(Y_i+\x3f_3)\dd43=\\
-2m_1^2\x3\dd33+(\x3-\x2)\cc42124+(\x2-\x1)\cc42134\\
+\x1\cc42234+\x3\cc32234 \nonumber
\end{multline}

\end{document}